\documentclass[a4paper, amsfonts, amssymb, amsmath, reprint, showkeys, nofootinbib, twoside, aps]{revtex4-1}

\usepackage{siunitx}
\usepackage{amsmath}
\usepackage{graphicx}
\usepackage{hyperref}
\usepackage{comment}
\DeclareUnicodeCharacter{2212}{\textminus}

\DeclareSIUnit\angstrom{\text {Å}}

\begin{document}

\title{A statistical understanding of oxygen vacancies in distorted high-entropy perovskite oxides}
\author{
Adam Potter\textsuperscript{1},
Yifan Wang\textsuperscript{1,2},
Kiran Hamkins\textsuperscript{1}, 
Dongjae Kong\textsuperscript{1},
Yuzhe Li\textsuperscript{2},
Jian Qin\textsuperscript{3},
Xiaolin Zheng\textsuperscript{1,4,*}
}

\affiliation{\textsuperscript{1}Department of Mechanical Engineering, Stanford University, Stanford, CA, USA}
\affiliation{\textsuperscript{2}Department of Material Science and Engineering, Stanford University, Stanford, CA, USA}
\affiliation{\textsuperscript{3}Department of Chemical Engineering, Stanford University, Stanford, CA, USA}
\affiliation{\textsuperscript{4}Department of Energy Science and Engineering, Stanford University, Stanford, CA, USA}
\thanks{Corresponding author: xlzheng@stanford.edu}

\begin{abstract}
High-entropy perovskite oxides have emerged as promising electrode materials for solid oxide electrolyzers. However, their compositional complexity makes the formation of oxygen vacancies, which influence properties such as oxygen ionic conductivity and thermal expansion, challenging to predict. 
Here, we experimentally measure changes in oxygen vacancy concentration for fourteen perovskite oxides with high and low-entropy A-site compositions, finding a dependence on cation size variance in addition to divalent cation fraction. 
Atomistic simulations using a machine-learned universal interatomic potential reveal cation size mismatches broaden a distribution of vacancy formation energies, shown through statistical thermodynamics to shift bulk formation thermodynamics. 
Treating oxygen vacancies statistically enables accurate predictions of oxygen vacancy formation compared to traditional models.
Practically, increasing the size variance between A-site cations reduces the temperature sensitivity of oxygen vacancy concentrations, making it key for tuning critical properties. More broadly, this study demonstrates statistical treatment of oxygen vacancies is essential for understanding high-entropy perovskite oxides.

\end{abstract}

\keywords{high-entropy perovskite oxide, oxygen vacancy, SOEC, oxygen electrode}

\maketitle

\section{Introduction}
%
Perovskite oxides (ABO$_{\rm 3-\delta}$), such as LSCF ($\rm La_{0.6} Sr_{ 0.4} Co_{0.2} Fe_{0.8} O_{3-\delta}$), are widely utilized as anodes for high-temperature solid oxide electrolyzer cells (SOECs), which offer the potential for efficient hydrogen production using renewable energy and heat. However, the performance of SOECs is often limited by anode properties, including low oxygen ion conductivity and high thermal expansion~\cite{pikalova_high-entropy_2022}. The oxygen ion conductivity of perovskites relies on a high concentration of oxygen vacancies ($\delta$) to enable efficient ion conduction via vacancy hopping \cite{xiang_high-entropy_2021}. The concentration of oxygen vacancies also affects other important properties, such as the thermal expansion coefficient, thermochemical stability, and electrical conductivity \cite{bae_defect_2020, choi_thermodynamic_2012}. As a result, understanding and careful control of oxygen vacancy behavior is critical for optimizing the performance of perovskite oxides in high-temperature SOECs. 

Recently, high entropy perovskite oxides (HEPOs) (ABO$_{\rm 3-\delta}$), with five or more cations with nearly equal molar amounts in either A- or B-site~\cite{rost_entropy-stabilized_2015}, have attracted great attention for SOECs as they theoretically could have good high-temperature chemical and structural stability due to the high mixing entropy~\cite{xiang_high-entropy_2021}. In addition, high-entropy perovskites were shown to introduce bond distortion~\cite{su_direct_2022} and local ordering~\cite{xu_local_2024}, which could potentially influence the formation of oxygen vacancies. It has been demonstrated that substituting metal cations can change the oxygen vacancy concentration \cite{zhang_tuning_2023, choi_oxygen_2012, bae_defect_2020, choi_oxygen_2012}.
However, exactly how the oxygen vacancy concentration in HEPOs varies with metal cation composition and temperature remains elusive as the net energy needed to remove a lattice oxygen is significantly complicated by variations from different metal cations and configurational entropy.  

Previous research on simple perovskites has provided valuable methodologies and insights for studying oxygen vacancy. For example, experimental studies vary concentrations of one or two metal cations systematically and measure oxygen vacancies using techniques, such as thermo-gravimetry analysis (TGA) \cite{oishi_oxygen_2008, luo_predicting_2014, choi_thermodynamic_2012, bae_investigations_2019, park_accurate_2023, mizusaki_nonstoichiometry_1985}, Columbic titration \cite{jeon_oxygen_2012, choi_correlation_2014, karppinen_oxygen_2002, choi_oxygen_2012}, X-ray photoelectron spectroscopy (XPS) \cite{zhang_tuning_2023}, or iodometric titration \cite{karppinen_oxygen_2002}. Then, the vacancies' formation thermodynamics can be determined by applying concentration data to established defect models \cite{bae_defect_2020, choi_correlation_2014}. Those studies consistently reveal the importance of the concentration of 2+ cations in the A-site ($X_{2+}$) for forming high oxygen vacancy concentrations \cite{bae_investigations_2019, choi_correlation_2014}. However, experimentally sweeping every metal cation concentration is impractical for high-entropy materials with five or more dopants, and no such experimental studies have been reported in the literature.

In comparison, there are a few computational studies that aim to predict high-entropy doping effects on the oxygen vacancy concentration by calculating the vacancy formation energy ($E_{\rm v}$). Density functional theory (DFT) studies calculated $E_{\rm v}$ for simple \cite{luo_predicting_2014} and high-entropy oxides \cite{park_accurate_2023, xu_local_2024, zhang_tuning_2023} and found cation substitutions could tune $E_{\rm v}$ as well as where vacancies preferred to form. An empirical study found such DFT-calculated $E_{\rm v}$ values could be predicted from the oxide's formation enthalpy, electronegativity, and band gap ~\cite{deml_intrinsic_2015}. However, while it is possible to predict $E_{\rm v}$ for many high-entropy oxide compositions computationally, these $E_{\rm v} $ predictions have yet to consistently align with experimentally measured oxygen vacancy concentrations. A recent study by Park et al.~\cite{park_accurate_2023} calculated $E_{\rm v} $ with DFT across many oxygen sites for one HEPO and three simpler oxides and found all four had significant deviations in $E_{\rm v}$ depending on their cation neighbors. To address this variability, Park proposed $E_{\rm v}$ values exist in a distribution, noting the potential implications of the mean and variance on equilibrium vacancy concentration. Xu et al.~\cite{xu_local_2024} also used a statistical approach in their DFT study of complex perovskites to predict oxygen vacancy formation, finding that vacancies preferentially form near cobalt sites. Both studies suggest the statistics of $E_{\rm v}$ may be key to predicting and tuning oxygen vacancies, however, the importance of such statistical effects for predictive models is unclear. The mechanisms of $E_{\rm v}$ statistics has also remained largely unexplored for a wide range of high-entropy oxides compositions, both experimentally and computationally. This gap needs to be filled to provide principles or guiding models for the selection of cation dopants to tune oxygen vacancies, and the critical properties dependent on them, in complex perovskite oxides.

In this work, we establish a combined experimental, computational, and theoretical workflow to study the effect of high-entropy mixed A-sites in perovskite oxides on the concentration of oxygen vacancies. Experimentally, we measured the oxygen vacancy concentration of 11 high-entropy (HE) and 3 low-entropy (LE) perovskites as a function of temperature using TGA, and found an enthalpy-entropy compensation relation where HE samples exhibit both lower enthalpy and entropy of oxygen vacancy formation. We found that the oxygen vacancy concentrations \textcolor{black}{in LSCF-based perovskite oxides} are directly affected by two primary variables: the conventional 2+ cation molar ratio in the A-site ($X_{2+}$), and the newly identified A-site ionic radius variance ($\sigma_A$).
Computationally, our high-throughput simulations with a machine-learned interatomic potential show that increased $\sigma_A$ distorts oxygen–B-site bonds, increasing the variance in oxygen bonding energy ($\sigma_{\rm Ev}$). Our statistical thermodynamic analysis shows that this broadening in $\sigma_{\rm Ev}$ reduces the vacancy enthalpy ($\Delta H^f$) and entropy ($\Delta S^f$) of formation, matching experimental trends. We further propose a statistical model of vacancy formation that accurately predicts the vacancy concentrations of LE and HE materials from simulation.
Overall, our results highlight A-site size mismatches in the A-site ($\sigma_A$) as a practical lever for tuning oxygen vacancies and their temperature dependence in LSCF-based perovskite oxides. Theoretically, tuning $\sigma_{\rm Ev}$ via structural distortion or other means could have similar effects in other oxide systems, high-entropy or not.

\begin{table*}[!ht]
\centering
\scalebox{0.9}{

\begin{tabular}{@{}|l|l|l|l|l|l|l|l|l|@{}}

\hline
\textbf{Label} & 
\textbf{A-Site Composition} & 
\begin{tabular}[c]{@{}l@{}}$X_{2+}$\\ \end{tabular} & 
\begin{tabular}[c]{@{}l@{}}$t$ \\ \end{tabular} & 
\begin{tabular}[c]{@{}l@{}} $\sigma_A$ \\  (\%)\end{tabular} & 
\begin{tabular}[c]{@{}l@{}} $\Delta S_{mix}$\\ (J mol$^{-1}$ K$^{-1}$)\end{tabular} & 
\begin{tabular}[c]{@{}l@{}}$\Delta H^f$\\ (kJ mol$^{-1}$)\end{tabular} &  
\begin{tabular}[c]{@{}l@{}}$\Delta S^f$\\ (J mol$^{-1}$ K$^{-1}$)\end{tabular} &  
\begin{tabular}[c]{@{}l@{}}$\delta_0$\\ ($10^{-3}$)\end{tabular} \\ 

\hline
LE-1           & La\textsubscript{0.6}Sr\textsubscript{0.4} (LSCF)                  & 0.40    & 0.978       & 2.82    & 5.60   & 90 $\pm$ 8     & 56 $\pm$ 7         & 0.79 $\pm$ 0.4   \\

LE-2           & La\textsubscript{0.6}Sr\textsubscript{0.2}Ca\textsubscript{0.2}   & 0.40        & 0.971     & 2.54     & 7.90   & 92 $\pm$ 8     & 60 $\pm$ 8     & 0.83 $\pm$ 0.4    \\

LE-3           & La\textsubscript{0.7}Ca\textsubscript{0.3}      & 0.30     & 0.965          & 0.68    & 5.08   & 77 $\pm$ 4     & 41 $\pm$ 3     & 0.55 $\pm$ 0.1     \\

HE-1           & La\textsubscript{0.17}Sr\textsubscript{0.17}Ca\textsubscript{0.17}Ba\textsubscript{0.17}Nd\textsubscript{0.17}Sm\textsubscript{0.17}    & 0.50    & 0.973     & 8.90    & 14.90  & 54 $\pm$ 7   & 33 $\pm$ 5         & 18 $\pm$ 9   \\

HE-2           & La\textsubscript{0.2}Sr\textsubscript{0.2}Ca\textsubscript{0.2}Nd\textsubscript{0.2}Y\textsubscript{0.2}    & 0.40   & 0.957  & 5.10    & 13.38   & 72 $\pm$ 4    & 57 $\pm$ 3     & 11 $\pm$ 3   \\

HE-3           & La\textsubscript{0.2}Sr\textsubscript{0.2}Ca\textsubscript{0.2}Nd\textsubscript{0.2}Sm\textsubscript{0.2}   & 0.40   & 0.957    & 5.10     & 13.38  & 71  $\pm$ 2    & 51 $\pm$ 2    & 7.6 $\pm$ 1     \\

HE-4           & La\textsubscript{0.2}Sr\textsubscript{0.2}Ca\textsubscript{0.2}Gd\textsubscript{0.2}Sm\textsubscript{0.2}   & 0.40   & 0.957    & 5.22   & 13.38 & 71 $\pm$ 2    & 52 $\pm$ 2   & 8.6 $\pm$ 1    \\

HE-5           & La\textsubscript{0.2}Sr\textsubscript{0.2}Ba\textsubscript{0.2}Gd\textsubscript{0.2}Nd\textsubscript{0.2}      & 0.40   & 0.978     & 9.05  & 13.38  & 52 $\pm$ 3     & 33 $\pm$ 2      & 14 $\pm$ 3  \\

HE-6           & La\textsubscript{0.4}Sr\textsubscript{0.2}Ba\textsubscript{0.1}Ca\textsubscript{0.1}Nd\textsubscript{0.2}    & 0.40      & 0.975   & 6.77    & 12.22  & 59 $\pm$ 1   & 35 $\pm$ 1   & 6.8 $\pm$ 0.3  \\

HE-7           & La\textsubscript{0.2}Ba\textsubscript{0.1}Ca\textsubscript{0.3}Nd\textsubscript{0.2}Y\textsubscript{0.2}    & 0.40    & 0.960    & 7.46   & 12.99  & 49 $\pm$ 5    & 26 $\pm$ 4     & 10 $\pm$ 4    \\

HE-8           & La\textsubscript{0.17}Sr\textsubscript{0.17}Ca\textsubscript{0.17}Gd\textsubscript{0.17}Nd\textsubscript{0.17}Sm\textsubscript{0.17}      & 0.33      & 0.953    & 5.10    & 14.90   & 72 $\pm$ 3     & 53 $\pm$ 3         & 5.2 $\pm$ 1   \\

HE-9           & La\textsubscript{0.14}Sr\textsubscript{0.14}Ca\textsubscript{0.14}Gd\textsubscript{0.14}Nd\textsubscript{0.14}Sm\textsubscript{0.14}Y\textsubscript{0.14} & 0.29     & 0.950  & 5.12   & 16.18  & 73 $\pm$ 8          & 53 $\pm$ 7     & 3.3 $\pm$ 2   \\

HE-10          & La\textsubscript{0.2}Sr\textsubscript{0.2}Gd\textsubscript{0.2}Sm\textsubscript{0.2}Y\textsubscript{0.2}    & 0.20   & 0.950     & 5.80    & 13.38    & 74 $\pm$ 11          & 53 $\pm$ 10     & 1.6 $\pm$ 1   \\

HE-11          & La\textsubscript{0.17}Sr\textsubscript{0.17}Gd\textsubscript{0.17}Nd\textsubscript{0.17}Sm\textsubscript{0.17}Y\textsubscript{0.17}       & 0.17    & 0.948    & 5.46    & 14.90  & 72 $\pm$ 5     & 51 $\pm$ 4    & 1.2 $\pm$ 0.4    \\ 

\hline
\end{tabular}
}
\caption{Low entropy (LE) and high entropy (HE) A-site compositions with B-Site (Co\textsubscript{0.2}Fe\textsubscript{0.8}) described with pre-calculated values: divalent mole fraction ($X_{2+}$), tolerance factor ($t$), size variance ($\sigma_A$), and mixing entropy ($\Delta S_{mix}$). Experimental formation enthalpies ($\Delta H^f$), entropies ($\Delta S^f$), initial vacancies at 500$^\circ C$ ($\delta_0$) and standard error fit from TGA data with Eq.\ref{eq:trad_keq_gibbs}. }
\label{table:sample_descriptors}
\end{table*}

\section{Results}
We study perovskites with LE (2-3 cations) and HE (5+ cations) A-sites taking the form A(Co$_{0.2}$Fe$_{0.8}$)O$_{3-\delta}$, derived from LSCF,  the state of the art for SOEC anodes \cite{pikalova_high-entropy_2022, jiang_development_2019}. A-sites consist of Lanthanides and Alkaline earth metals and span a set of common compositional descriptors, such as divalent metal cation concentration $X_{2+}$ (Eq. \ref{eq:X_2_plus}), A-site cation size variance $\sigma_A$ (Eq. \ref{eq:size_variance}), Goldschmidt tolerance factor $t$ (Eq. \ref{eq:tolerance_factor}),  and mixing entropy $\Delta S_{mix}$ (Eq. \ref{eq:entropy_mixing}). We measure samples with both different and same A-site divalent (2+ cation) doping concentrations ($X_{2+}$) to differentiate divalent doping from other A-site mixing effects.

\subsection{Experimental Estimation of Oxygen Vacancy Formation }

\begin{figure}[!ht]
    \centering
    \includegraphics[width=1.0\linewidth]{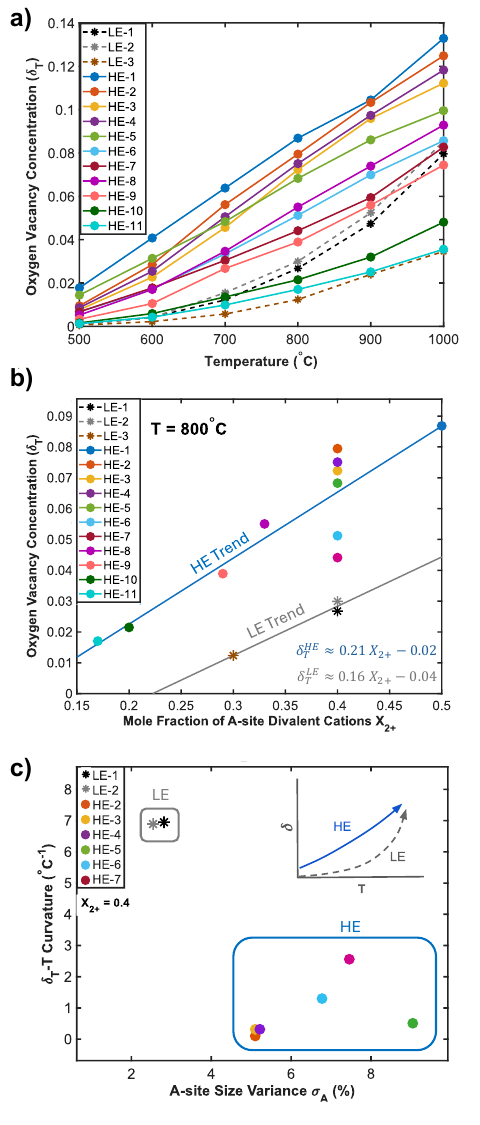}
    
    \caption{ \textbf{Oxygen vacancy formation curves. a} Experimental oxygen vacancy concentration ($\delta_T$) as a function of temperature from TGA for low-entropy (LE) and high-entropy (HE) compositions. \textbf{b} vacancy concentration at 800$^\circ C$ ($\delta_{800^\circ C}$) as a function of A-site 2+ site fraction ($X_{2+}$). \textbf{c} Curvature of $\delta_T$-T curves in Figure \ref{fig:tga_panel}a as a function A-site size variance ($\sigma_A$) for samples with $X_{2+}$=0.4 and (top right) drawn schematic of differences in vacancy formation. Source data are provided as a Source Data file.}
    \label{fig:tga_panel}
\end{figure}

All 14 perovskite oxides were synthesized using solution combustion synthesis followed by thermal annealing at 800$^\circ$C and those oxide powders were further processed with high-speed ball-milling to break apart agglomerations (see Methods). Those oxides show a homogeneous distribution of elements, as illustrated by a representative sample HE-2 using scanning electron microscopy (SEM) paired with energy dispersive X-rays (EDX) (Supplementary Figure 1a). The structure of all 14 oxides was characterized with X-ray diffraction (XRD) and found to match the same perovskite phase (orthorhombic Pnma) (Supplementary Figure 1b). While all synthesized HE oxides in the study were single phase, predicting stable high-entropy compositions remains a challenge for other compositions \cite{sivak_discovering_2025, pitike_predicting_2020, sarker_high-entropy_2018}.

The change in oxygen vacancy concentration at temperature $T$, $\Delta\delta_T$, was measured using TGA. The weight was stabilized at 300$^\circ$C  under ambient air, then increased from 500 to 1000$^\circ$C with 1-2 hour holds every 100$^\circ$C. Stabilized mass readings suggest the vacancy formation reaction was in equilibrium (Supplementary Figure 2). Changes in oxygen vacancy concentration, $\Delta\delta_T$, are offset by $\delta_0$  ($\delta_T= \delta_0 + \Delta\delta_T$, where $\delta_0 = \delta(500^\circ  \rm C)$) to find vacancy concentrations $\delta_T$. The oxygen vacancy concentrations are shown in Figure \ref{fig:tga_panel}a for all samples. Values of $\delta_0$ are expected to be small and were fit using nonlinear least squares (Table \ref{table:sample_descriptors}) to a reaction model described below (Eq. \ref{eq:trad_keq_gibbs}), a typical approach for similar studies \cite{oishi_oxygen_2008, choi_thermodynamic_2012, bae_investigations_2019}. The fitted $\delta_0$ values range from 0.0005-0.014 and are indeed small compared to the measured concentration increase ranging from 0.035-0.13 seen in Figure \ref{fig:tga_panel}a. As such, errors in model-fit $\delta_0$ values are likely to have marginal effects on the overall vacancy formation trends. The measured $\delta(T)$ curve for LE-1 (equivalent to LSCF) in Figure \ref{fig:tga_panel}a is in agreement with reported values for the similar composition (La\textsubscript{0.6}Sr\textsubscript{0.4}Fe)O\textsubscript{3}: $\delta$=0.044 vs 0.035 in \cite{kuhn_oxygen_2011} at 900$^\circ  \rm C$.

Upon close examination of Figure \ref{fig:tga_panel}a, it can be seen that the HE samples (solid lines) tend to exhibit more oxygen vacancies than LE samples (dashed lines), particularly at temperatures $\leq$ 900$^\circ C$. Furthermore, the curves of the HE samples are quasilinear, but the LE samples show more exponential growth of oxygen vacancy concentration with temperature. Even among HE samples, HE-3 and HE-4 have steeper curves than HE-7 or HE-5. Figure \ref{fig:tga_panel}b shows the oxygen vacancy concentration at 800$^\circ C$ as a function of the mole fraction of 2+ cations in the A-site ($X_{2+}$), and clearly, larger values of $X_{2+}$ lead to more vacancies in general for LE and HE samples. This is because $X_{2+}$ is directly tied to the formation of oxygen vacancies as explained in Section \ref{sec:defect_modeling}. Despite the important role of $X_{2+}$, there are still large unexplained variations in the concentration of oxygen vacancy between samples for the same $X_{2+}$=0.4, indicating that other factors also play an important role. Figure \ref{fig:tga_panel}c plots the curvature of Figure \ref{fig:tga_panel}a (in units $^\circ C^{-1}$) against the A-site size variance $\sigma_A$. The HE samples have consistently smaller curvatures and thus more linear growth of oxygen vacancies. $\sigma_A$ is given by Eq. \ref{eq:size_variance} in percent where $r_{A, i}$ is the ionic radius of the $i$th A-site element out of $N$ elements and $\hat{r}_A$ is the average radius. 

\begin{equation}
    \sigma_A = 100 \cdot \frac{\sqrt{ \sum_i^N (r_{A,i}-\hat{r}_A)^2 }}{\sum_i^N r_A,i} 
    \label{eq:size_variance}
\end{equation}

The importance of $X_{2+}$ is well known in the literature and will be discussed in the context of defect reaction thermodynamics in the following section. In comparison, the importance of $\sigma_A$ in oxygen vacancy thermodynamics is yet to be explored and will be the subject for the remainder of this paper.

\subsection{Oxygen Vacancy Formation Thermodynamics}
\label{sec:defect_modeling}
This section aims to determine the $\Delta G^f$, $\Delta H^f$, and $\Delta S^f$ values for the oxygen formation reaction (Eq. \ref{eq:vac_form_rxn2}) for both the LE and HE samples. In the reaction given by Eq. \ref{eq:vac_form_rxn2}, $O^x$ is lattice oxygen, $h^.$ are holes, and $V^{..}_O$ are oxygen vacancies in Kroger-Vink notation. When the A-site of a perovskite is doped with a 2+ cation ($X_{2+}$), its negative charge relative to other 3+ cations will be compensated by forming either a B-site hole ($h^{.}$) or half of an oxygen vacancy ($V^{..}_O$) to maintain charge neutrality as expressed by Eq.\ref{eq:charge_neutrality}. The holes can be thought of as 4+ oxidation states of B-site cations (Fe$^.$ or Co$^.$) with a net positive defect charge relative to the typical 3+ state, as the 2+ oxidation state of B-site elements (Fe$^{'}$, Co$^{'}$) has negligible concentrations in oxidizing conditions \cite{choi_correlation_2014, oishi_oxygen_2008}. So long as Eq. \ref{eq:charge_neutrality} is relevant, the maximum concentration of oxygen vacancies is limited by hole charge availability such that:  $   \rm{max}([{\rm V}^{..}_{\rm O}]) =  \frac{1}{2}[{\rm X}_{2+}^{'}]$. 
This relation assumes the following: 1) 2+ A-site doping is the dominant form of hole generation (i.e., the intrinsic charge carrier concentration is negligible); 2) only the B-site cations change oxidation states during the oxygen vacancy formation; and 3) holes are not differentiated by B-site element  \cite{choi_correlation_2014, kuhn_oxygen_2011, sogaard_oxygen_2007}.
Moreover, as a perovskite unit cell has three oxygen sites, the sum of vacancies $V^{..}_O$ and occupied sites $O^x_O$ must equal three (Eq.\ref{eq:oxygen_conservation}). 

\begin{equation}
    {\rm O}^x_{\rm O} + 2{\rm h}^.
    \leftrightarrow {\rm V}^{..}_{\rm O} +
    \frac{1}{2} {\rm O}_2 
    \label{eq:vac_form_rxn2}
\end{equation}
\begin{equation}
    [{\rm X}_{2+}^{'}] = 2[{\rm V}^{..}_{\rm O}] + [{\rm h}^{.}]
    \label{eq:charge_neutrality}
\end{equation}
\begin{equation}
    [{\rm O}^x_{\rm O}] + [{\rm V}^{..}_{\rm O}] = 3
    \label{eq:oxygen_conservation}
\end{equation}

The equilibrium constant, $K_{\rm p,ox}$ for the oxygen formation reaction (Eq. \ref{eq:vac_form_rxn2}) can be expressed as Eq. \ref{eq:Keq_derived} considering the concentration relations in Eqs. \ref{eq:charge_neutrality}-\ref{eq:oxygen_conservation}  and assuming a reference oxygen partial pressure ($pO_2^0$) of 1 bar. $K_{\rm p,ox}$ is now solely a function of $X_{2+}$, $pO_2$, and $\delta_T$ (Eq. \ref{eq:Keq_derived}) which are known or measured with TGA (Figure \ref{fig:tga_panel}). Similar or identical expressions for the oxygen vacancy equilibrium constant have been used in previous studies \cite{oishi_oxygen_2008, choi_thermodynamic_2012, bae_investigations_2019, mizusaki_nonstoichiometry_1985}. The explicit relation of oxygen vacancies, $\delta_T$, on the concentration of $X_{2+}$ in this expression is consistent with that seen in Figure \ref{fig:tga_panel}b. $X_{2+}$ is critical because previous studies have shown oxygen vacancy concentration saturates at $\delta = \frac{1}{2}X_{2+}$ due to polaron depletion until the system reaches highly reducing conditions (${\rm pO}_2 \leq 10^{-10}$ atm for similar materials) \cite{sogaard_oxygen_2007, bae_investigations_2019, oishi_oxygen_2008}.

\begin{equation}
    K_{\rm p,ox} (T) = 
    \frac{[{\rm V}^{..}_{\rm O}]
    \left[\frac{p{\rm O_2}}{p{\rm O_2^0}}\right]^{1/2}}
    {[{\rm O}^x_{\rm O}][{\rm h}^.]^2} =
    \frac{\delta_T \sqrt{p{\rm O_2}}}{(3-\delta_T)
    ([{\rm X}_{2+}]-2\delta_T)^2} 
    \label{eq:Keq_derived}
\end{equation}

This equilibrium constant can also be expressed as a function of the Gibbs free energy of formation $\Delta G^f$, or equivalently enthalpy of formation $\Delta H^f$ and entropy of formation $\Delta S^f$ of oxygen vacancy, the specific gas constant $R$, and temperature $T$ (Eq. \ref{eq:vant_hoff} ). The combination of Eqs. \ref{eq:Keq_derived}, and \ref{eq:vant_hoff} result in Eq. \ref{eq:trad_keq_gibbs} below. 
\begin{equation}
    K_{\rm p,ox}(T) = e^{-\frac{\Delta G^f}{RT}} = e^{\frac{\Delta S^f}{R}-\frac{\Delta H^f}{RT}}
    \label{eq:vant_hoff}
\end{equation}
\begin{equation}
\begin{split}
   & \frac{\Delta S^f}{R}-\frac{\Delta H^f}{RT}
   =  \\
   & \ln\left[ {\frac{(\delta_0 + \Delta\delta_T) 
   \sqrt{p{\rm O_2}}}{(3-(\delta_0 + \Delta\delta_T))([{\rm X}_{2+}]-
   2(\delta_0 + \Delta\delta_T))^2} } \right]
   \end{split}
    \label{eq:trad_keq_gibbs}
\end{equation}

Now, the $\delta_T$ curves measured by TGA for all 14 samples (Figure 2a) can be fit to Eq. \ref{eq:trad_keq_gibbs} using nonlinear least squares to extract $\Delta H^f$, $\Delta S^f$, and $\delta_0$ ($R^2>$0.99 for all samples) and the values are reported in Table \ref{table:sample_descriptors}. The fitted values of $\Delta H^f$ and $\Delta S^f$ for LE-1 are similar to previously reported values for the similar oxide $\rm (La_{0.6}Sr_{0.4}FeO_3)$: 90 kJ mol$^{-1}$ vs 95 kJ mol$^{-1}$ \cite{kuhn_oxygen_2011} and 90 kJ mol$^{-1}$ \cite{sogaard_oxygen_2007} for $\Delta H^f$ and 56 J mol$^{-1}$ K$^{-1}$ vs 54 J mol$^{-1}$ K$^{-1}$ \cite{kuhn_oxygen_2011} and 58 J mol$^{-1}$ K$^{-1}$ \cite{sogaard_oxygen_2007} for $\Delta S^f$, validating the methodology. 

Notably, the model in Eq. \ref{eq:Keq_derived} is only valid for oxides with extrinsic polaron holes that interact with oxygen vacancies, as is widely accepted for LSCF-like perovskite oxides \cite{jiang_development_2019}. This occurs when the divalent doping exceeds oxygen vacancy concentrations ($\delta \leq \frac{1}{2} X_{2+}$) which is typical in the oxidizing conditions used in this study (${\rm pO}_2 = 0.21$ atm). Furthermore, this model does not differentiate polarons by B-site, in part because polarons prefer forming on iron over cobalt \cite{jiang_development_2019, hartmann_investigation_2023} so long as iron sites are available. For lower iron versus cobalt compositions where $[{\rm Fe}] < X_{2+}$ polarons could overflow onto cobalt sites affecting oxygen vacancy dynamics.

\subsection{Differences in Vacancy Formation for HE and LE Samples}
\label{sec:params_exp}

\begin{figure*}[!ht]
    \centering
    \includegraphics[width=0.85\linewidth]{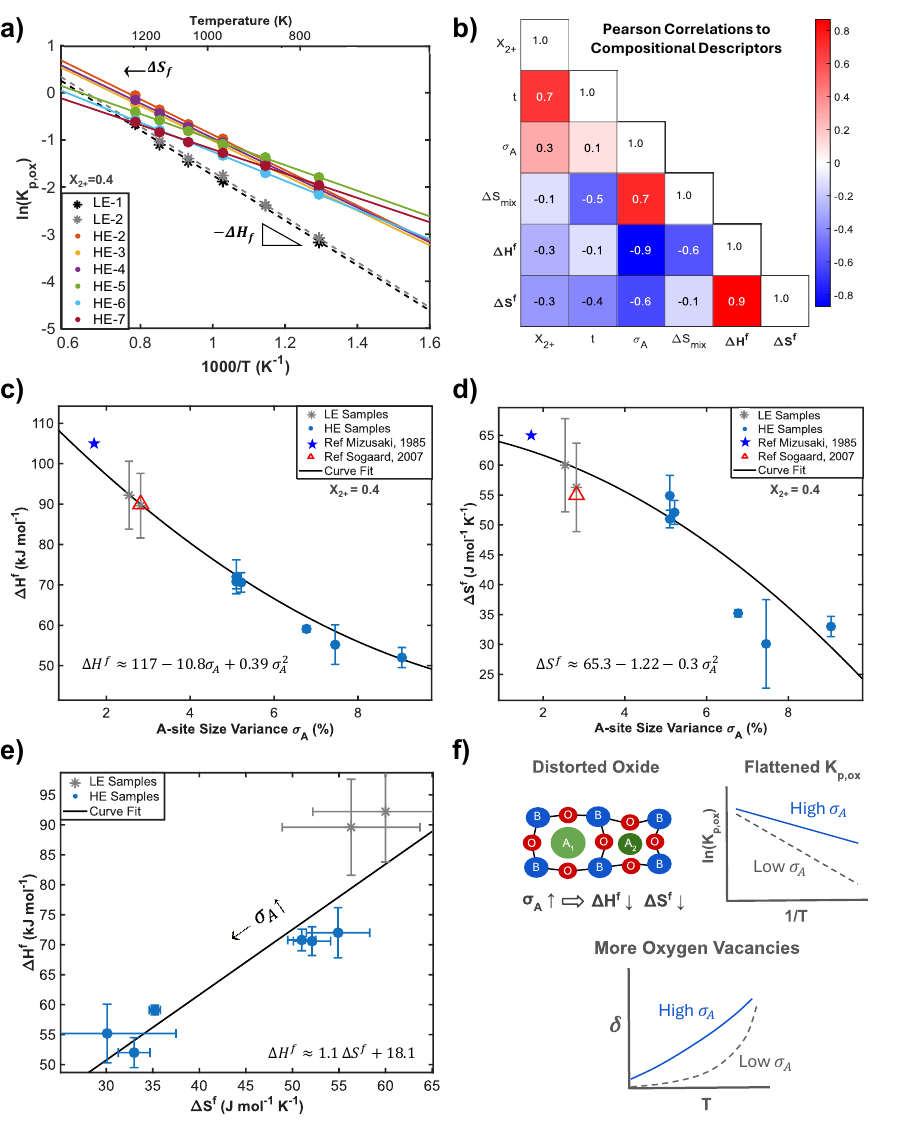}
    \caption{\textbf{Composition trends and thermodynamics. a} Van't Hoff diagram showing the oxygen vacancy formation equilibrium constant ($K_{\rm p,ox}$) for samples with the same 2+ A-site fraction, $X_{2+}=0.4$. \textbf{b} Pearson correlation matrix linking oxygen vacancy formation enthalpy ($\Delta H^f$) and entropy ($\Delta S^f$) to four compositional descriptors. \textbf{c-d}  $\Delta H^f$ and $\Delta S^f$ as a function of A-site size variance ($\sigma_{A}$) with standard errors and values from \cite{mizusaki_nonstoichiometry_1985} (blue star) and \cite{sogaard_oxygen_2007} (red triangle). \textbf{e} Correlation between $\Delta H^f$ and $\Delta S^f$ roughly scaling with $\sigma_A$.  \textbf{f}  Drawn schematic of a generic perovskite oxide (green A-, blue B-, and red O-sites) with a high $\sigma_{A}$ due to $A_1$ being larger than $A_2$. This leads to a smaller $\Delta H^f$, $\Delta S^f$, and flatter $K_{\rm p,ox}$ corresponding to higher oxygen vacancy concentrations that grow linearly with temperature. Source data are provided as a Source Data file.}
    \label{fig:empirical_panel}
\end{figure*}

This section aims to understand differences in $\Delta G^f(T)$ between HE and LE samples that lead to the oxygen vacancy concentration differences at various temperatures. The Van't Hoff diagram (Figure \ref{fig:empirical_panel}a) shows $\ln K_{\rm p,ox}$ (or $-\Delta G^f/RT$) vs $1/T$ (Eq. \ref{eq:vant_hoff}) exclusively for samples with same $X_{2+}=0.4$. $\Delta S^f$  and $\Delta H^f$ and are the intercept and negative of the slope, respectively. LE and HE samples show different trends; HE samples exhibit a larger $K_{\rm p,ox}$ (smaller $\Delta G^f$) at lower temperatures, indicating more oxygen vacancies. With increasing temperature, the $K_{\rm p,ox}$ of HE samples increases more slowly than that of LE samples due to their smaller $\Delta H^f$. This is equivalent to the oxygen vacancy concentration of HE samples increasing linearly, instead of exponentially, with temperature (Figure \ref{fig:tga_panel}a and \ref{fig:tga_panel}c). 

We conducted a correlation study between the measured $\Delta H^f$ and $\Delta S^f$  (Table \ref{table:sample_descriptors}) and four compositional descriptors ($X_{2+}$, $t$, $\sigma_A$, $\Delta S_{mix}$) across n=14 samples (see details in Methods) shown in Figure \ref{fig:empirical_panel}b. Such a correlation study has not been reported previously. First, the cross-correlation of the four compositional descriptors is limited, suggesting the chosen sample compositions effectively span a four-dimensional A-site design space. In addition, $\Delta H^f$ and $\Delta S^f$ are not strongly correlated with $X_{2+}$, indicating the dependence seen in Figure \ref{fig:tga_panel}b is encapsulated by the inclusion of $X_{2+}$ in the equilibrium constant (Eq. \ref{eq:Keq_derived}). $\Delta H^f$ and $\Delta S^f$ show moderate correlations with $\Delta S_{mix}$ and tolerance factor $t$, but strong correlation with the A-site size variance $\sigma_A$.

When our experimentally measured $\Delta H^f$ and $\Delta S^f$ (isolating $X_{2+}=0.4$) are plotted against $\sigma_A$ (Figures \ref{fig:empirical_panel}c and \ref{fig:empirical_panel}d), their clear correlations can be fit with empirical relations included in the plots. When we added $\Delta H^f$ and $\Delta S^f$ values reported by \cite{mizusaki_nonstoichiometry_1985} and \cite{sogaard_oxygen_2007} for variants of LE-1, they nicely follow the trends. The $\Delta H^f$ relationship with $\sigma_A$ in Figure \ref{fig:empirical_panel}c ($R^2$\qty{=0.95}) and  $\Delta S^f$ relationship in Figure \ref{fig:empirical_panel}d ($R^2$\qty{=0.83}) are quasilinear or quadratic. Together, we can observe a positive correlation between $\Delta H^f$ and $\Delta S^f$ in Figure~\ref{fig:empirical_panel}e.  This empirical correlation between $\Delta H^f$ and $\Delta S^f$ within a family of materials, known as enthalpy-entropy compensation, has been documented for reactions in other systems, although there is no agreement on the underlying mechanism~\cite{starikov_enthalpyentropy_2007, cornish-bowden_entropy-enthalpy_2018}. Suggested mechanisms include solvent interactions, hidden reaction intermediates \cite{starikov_enthalpyentropy_2007}, and numerical artifacts from data fitting \cite{pan_enthalpyentropy_2016}, the last of which should not apply as our $\Delta G^f$ values are distinct between LE and HE samples (Figure \ref{fig:empirical_panel}a).

A proposed relationship between oxygen vacancies and the A-site size variations is summarized in Figure \ref{fig:empirical_panel}f. Introducing mismatches in the A-site cation size induces distortion in the lattice \cite{su_direct_2022} and strains oxygen bonds \cite{park_accurate_2023}, which reduces $\Delta H^f$ and $\Delta S^f$. In short, a larger $\sigma_A$ (e.g. HE samples) leads to a flatter $\Delta G^f$, and hence a slower increase in oxygen vacancy with temperature (Figure \ref{fig:tga_panel}a). The next section dives into how tuning $\sigma_A$ can flatten $\Delta G^f$ by lowering $\Delta H^f$ and $\Delta S^f$. 

\subsection{Atomistic Simulations: Impact of $\sigma_A$ on Formation Energies} 
\label{sim_section}

The formation enthalpy $\Delta H^f$ of oxygen vacancies is a function of the energy to create an oxygen vacancy, $E_{\rm v} $, which can be calculated with atomistic simulations. 

\begin{equation}
    \Delta H^f = f(E_{\rm v} )
\end{equation}

A unique aspect of disordered high entropy materials is that the vacancy energies $E_{\rm v} $ vary across lattice sites, so we need to estimate $\Delta H^f$ and $\Delta S^f$ statistically using a distribution of vacancy energies $g(E_{\rm v} )$. Treating $E_{\rm v}$ statistically is unconventional and has only recently been considered \cite{park_accurate_2023,xu_local_2024}. We accomplished this by sampling hundreds of oxygen sites in an atomistic simulation using a machine learning universal interatomic potential (MLUIP) provided by Matlantis \cite{matlantis_paper}.  Included in the Supplementary Information is a detailed comparison of the Matlantis potential against DFT for calculating oxygen vacancy energy in smaller supercells over a selection of relevant materials (Supplementary Section 4). To control for contributions from divalent doping, only the eight materials in Table \ref{table:sample_descriptors} with $X_{2+}$= 0.4 are considered. All materials are modeled with an orthorhombic \textit{Pnma} perovskite phase, which fits well with XRD data (Supplementary Figure 1). Large randomly populated perovskite supercells are generated in the Atomic Simulation Environment \cite{ase} to represent LE and HE samples (Supplementary Figure 3). The resulting simulation cell contains 1280 atoms (256 perovskite unit cells), large enough that even the sparsest A-site element is represented within 2\% of its stated mole fraction.

\begin{figure*}[!ht]
    \centering
    \includegraphics[width=0.95\linewidth]{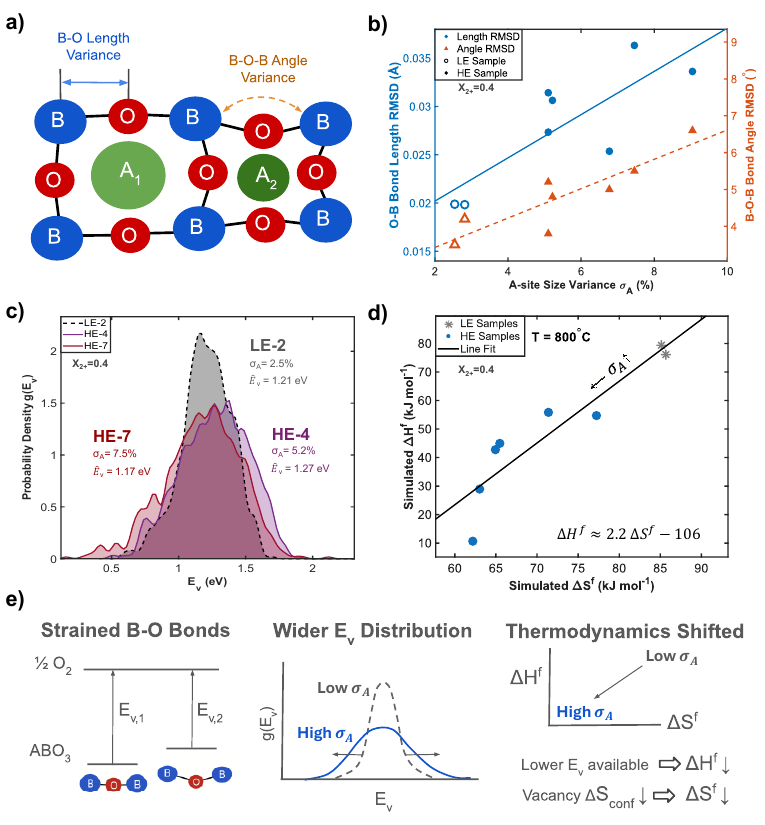}
    \caption{\textbf{Simulated structural distortion and vacancy energies. a} Schematic of generic perovskite (green A-, blue B-, and red O-sites) defining the length and angle of oxygen bonded to B-site cations. \textbf{b} Root mean squared deviation in bond length and angle as a function of A-site size variance ($\sigma_A$). \textbf{c} Simulation sampled distributions of vacancy formation energies ( $g(E_{\rm v} )$ ) for select LE and HE compositions. \textbf{d} Correlation between formation enthalpy ($\Delta H^f$) and entropy ($\Delta S^f$) predicted from simulated $g(E_{\rm v} )$ with Eqs. \ref{eq:H_approx}-\ref{eq:S_approx}. \textbf{e} Drawn schematics outlining how A-site size mismatch can strain B-O bonds, increasing variance in $E_{\rm v}$ resulting in both lower $\Delta H^f$ and $\Delta S^f$. Source data are provided as a Source Data file.}.
    \label{fig:sim_panel}
\end{figure*}

Vacancy energy $E_{\rm v} $ will be related to the B-site oxygen bonding as it is typically stronger than A-site oxygen bonding and dominates the bonding energy in perovskites  \cite{jing_role_2020}. Intuitively, variations in A-site size will cause variations in B-site bonding (Figure \ref{fig:sim_panel}a) \cite{jing_role_2020}. Such octahedra distortions have been directly observed with scanning transmission electron microscopy (STEM) in high-entropy oxides \cite{su_direct_2022,zou_high-entropy-induced_2025}. To quantify this, we first find the energy-minimized structures of the initial supercells of various HE and LE samples with no oxygen vacancies.  After energy minimization, we record the bond lengths and bond angles of B-O bonds for all 768 oxygen atoms (see Methods). Variance in bond length is known to affect properties in high-entropy materials and has been used as a predictor of phase stability \cite{sivak_discovering_2025}. As shown in Figure \ref{fig:sim_panel}b, a large variance in A-site cation size $\sigma_{\rm A}$ leads to higher root mean squared deviations (RMSD) for both the B-O bond lengths and the B-O-B bond angles. This suggests that a large $\sigma_{\rm A}$ leads to a broader distribution of B-site oxygen bonding geometries, hence a broader distribution of energy to create an oxygen vacancy. 

The oxygen vacancy formation energy $E_{\rm v} $ is calculated using Eq.~\ref{eq:E_sim_calc}, where $E_{\rm bulk}$ and $E_{\rm defect}$ are the total energy of a relaxed supercell before and after removing one oxygen atom,  and $ E_{\rm O_2}$  is the energy from forming gaseous oxygen found to be $-9.71$\,eV with the Matlantis potential (compared to the DFT tabulated elemental phase reference $-9.52$ eV \cite{stevanovic_correcting_2012}). It should be noted that our calculated vacancy formation energy values range from 0 to 2.5 eV, much larger than the mean average error of 0.03 eV relative to DFT the MLUIP reports for disordered systems  \cite{matlantis_paper}. 
\begin{equation}
    E_{\rm v}  = E_{\rm defect} - E_{\rm bulk} + \frac{1}{2} E_{\rm O_2}
    \label{eq:E_sim_calc}
\end{equation}
Calculated $E_{\rm v} $ values for HE samples have higher variance compared to LE samples (Table \ref{tab:sim_results}). $g(E_{\rm v} )$ represents the probability distribution of a given oxygen site having vacancy formation energy $E_{\rm v} $. Curves for $g(E_{\rm v} )$ in Figure \ref{fig:empirical_panel}c are estimated by sampling $E_{\rm v}$ for all oxygen sites in the supercell and smoothing the distribution using a kernel density estimation with bandwidth 0.03 eV. Figure \ref{fig:sim_panel}c shows that the means $\hat{E_{\rm v} }$ for the two HE and one LE samples are comparable, but the two HE samples have a higher variance, as we expected from Figure \ref{fig:sim_panel}b. The longer tail of the distribution, where $E_{\rm v} <$0.6eV suggests that more oxygen vacancies will be generated at lower temperatures for HE samples. \textit{Park et al.} suggested a similar effect of high-variance $g(E_{\rm v} $) distributions in their DFT study of disordered perovskites, although experimental verification of this model was inconclusive \cite{park_accurate_2023}. Notably, the average sampled vacancy energy $\hat{E_{\rm v} }$ across all HE samples was comparable the average for LE samples (1.19 vs 1.18 eV) while their average vacancy energy standard deviations $\sigma_{Ev}$ diverged significantly (0.19 vs 0.27 eV) suggesting $E_{\rm v} $ variance is key to differentiating LE and HE vacancy behavior (Table \ref{tab:sim_results}). Next, we will use statistical thermodynamics to understand how the broadening of $g(E_{\rm v} )$ in HE samples leads to observed differences in $\Delta H^f$ and $\Delta S^f$. 

\begin{table}[!ht]
\begin{tabular}{|l|l|l|l|l|}
\hline
\textbf{Label} & \textbf{$\hat{E}_{\rm v}$ }(eV) & \textbf{Avg $\hat{E}_{\rm v}$}(eV)& \textbf{$\sigma_{\rm Ev}$}(eV) & \textbf{Avg $\sigma_{\rm Ev}$}(eV)\\ \hline
LE-1           & 1.16                    & 1.18                  & 0.19&     0.19\\
LE-2           & 1.21                    &                  & 0.19&                                               \\ \hline
HE-2           & 1.27                    & 1.19                  & 0.28&     0.27\\
HE-3           & 1.29&                 & 0.28&                                              \\
HE-4           & 1.27                     &                 & 0.25&                                              \\
HE-5           & 1.00                     &                 & 0.29&                                              \\
HE-6           & 1.13                    &                  & 0.23&                                              \\
HE-7           & 1.17                     &                & 0.29&                                            \\ \hline
\end{tabular}
\caption{Simulated oxygen vacancy energy distributions ($g(E_{\rm v})$) with mean $\hat{E}_{\rm v}$ and standard deviation $\sigma_{\rm Ev}$ sampled from 768 oxygen sites for compositions with $X_{2+}$= 0.4 including average values over LE and HE samples}
\label{tab:sim_results}
\end{table}

\subsection{Oxygen Vacancy Energy and Statistical Thermodynamics}
\label{sec:theory}
Oxygen vacancies can be modeled as the exchange of lattice oxygen with an external reservoir at a fixed temperature, described by the grand canonical ensemble. Each site either binds to an atom or not, so they can be treated as ideal Fermions following Fermi-Dirac statistics. In this ensemble, the probability of site $i$ being vacant, $p_{{\rm v},i}$, is given by Eq.~\ref{eq:grand_canonical} where $\epsilon_i$ is the free energy of an oxygen vacancy on site $i$, $\mu_{\rm v}$ is the chemical potential of the oxygen vacancies in the system, and $k_{\rm B}$ is the Boltzmann constant. The oxygen vacancy fraction $x$, equivalent to the vacancy concentration $\delta$ over three, is the probability of vacancy averaged over $N$ oxygen sites (Eq.~\ref{eq:vac_from_prob}). 
\begin{equation}
    p_{{\rm v},i} =
     \frac{1}{1 + e^{(\epsilon_i - \mu_{\rm v})/ k_{\rm B} T}}  
    \label{eq:grand_canonical}
\end{equation} 
\begin{equation}
     x := \frac{\delta}{3} = \frac{1}{N} \sum_{i=1}^N p_{{\rm v},i}
    \label{eq:vac_from_prob}
\end{equation}
This vacancy probability $p_{{\rm v},i}$ is key to the statistical calculation of $\Delta H^f$ and $\Delta S^f$. Finding $p_{{\rm v},i}$ (Eq.~\ref{eq:grand_canonical}) depends on the vacancy free energy at each site $\epsilon_i$, which we will treat as a distribution with average free energy $\hat{\epsilon}$ and variance $\sigma_{\epsilon}^2$ (Eq. \ref{eq:Ev_variance}) due to small perturbations from the mean. 

\begin{equation}
    \sigma_{\epsilon}^2 = \frac{1}{N} \sum_{i=1}^{N} (\epsilon_i - \hat{\epsilon})^2
    \label{eq:Ev_variance}
\end{equation}
Given this problem setup, the objective of this derivation is to find the first-order effects of energy perturbations due to variance $\sigma^2_{\epsilon}$ on predicted values of $\Delta H^f$ and $\Delta S^f$. The first-order effects are derived in detail in the Supplementary Section 3. In short, by expanding around first-order terms of $\sigma^2_{\epsilon}$, we inverted Eqs. \ref{eq:grand_canonical}-\ref{eq:vac_from_prob} to arrive at an expression for oxygen vacancy chemical potential $\mu_{\rm v}$ as a function of vacancy fraction $x$ and $\sigma_{\epsilon}^2$ (Eq.~\ref{eq:chem_pot}). Eq.~\ref{eq:chem_pot} conforms to that for the regular solution theory \cite{rudisill_standard_1992}: the correction to the ideal mixing given by the rightmost term, of scale $\sigma_{\epsilon}^2/(k_{\rm B} T)$, is the first-order contribution of energy dispersion. When energy broadening $\sigma_{\epsilon}$ is sufficiently large, this additional term can be a substantial correction which has not been reported before and can be prominent for high entropy materials.
\begin{equation}
    \mu_{\rm v} = \hat{\epsilon} + k_{\rm B} T
    \ln\left(\frac{x}{1-x}\right) +
    \left(x-1/2\right)\frac{\sigma_{\epsilon}^2}{k_{\rm B} T}
    \label{eq:chem_pot}
\end{equation}

In equilibrium, the chemical potential of lattice oxygen equals that of oxygen in the environment. Constitutive assumptions outlined in the Supplementary Section 3 connect the free energy of vacancies $\epsilon_i$ and vacancy chemical potential $ \mu_{\rm v}$ to vacancy energy $E_{\rm v} $ and gaseous oxygen formation entropy $S_{\rm O_2}$ (a tabulated value). We assume that variations in free energy are primarily due to variations in the vacancy energy, that is, $\sigma_ {\epsilon}= \rm Var \rm[E_{\rm v} ] = \sigma_{\rm Ev}$. We define the distribution of oxygen vacancy energies with mean $\hat{E}_{\rm v}$ and variance $ \sigma^2_{\rm Ev}$ as $g(\hat{E}_{\rm v}, \sigma^2_{\rm Ev})$. Equating chemical potentials and collecting terms results in expressions for $\Delta H^f$ and $\Delta S^f$ as they are used in the law of mass action (Eqs. \ref{eq:Keq_derived}-\ref{eq:vant_hoff}). This results in Eqs. \ref{eq:H_approx} and \ref{eq:S_approx} which are valid for a small $ \sigma_{\rm Ev}$ and small vacancy fractions ($x $). Full details, including expressions valid for larger vacancy fractions ($x$),  can be found in the Supplementary. Notably, this derivation is applicable to oxygen vacancies in any material, and it has the potential to be extendable to other types of defects with variable formation energies. 

The first derived expression, Eq.~\ref{eq:H_approx}, predicts oxygen vacancy formation enthalpy $\Delta H^f$ including the first-order effects from broadening $g(\hat{E}_{\rm v}, \sigma^2_{\rm Ev})$. This expression is a departure from the assumption of uniform vacancies, where $\Delta H^f = \hat{E_{\rm v} }$ \cite{luo_predicting_2014,  zhang_tuning_2023, elmutasim_evolution_2024, park_accurate_2023} ,  and is relevant for disordered or high-entropy materials, where $\sigma_{\rm Ev} > 0$. This correction can be large: for $\sigma_{\rm Ev} = 0.2$eV, our typical value for HE materials, this correction reduces $\Delta H^f$ by 45 kJ mol$^{-1}$ at 1000K compared to experimental values of $\Delta H^f$=90 kJ mol$^{-1}$ for LE materials. At the limit of small vacancy concentrations, broadening $g(\hat{E}_{\rm v}, \sigma^2_{\rm Ev})$ will always be lower $\Delta H^f$. This is an intuitive result because $g(\hat{E}_{\rm v}, \sigma^2_{\rm Ev})$ will have a larger tail in the low $E_{\rm v} $ region; more low $E_{\rm v} $ oxygen sites will be available for vacancies. This is consistent with \textit{Park's} numerical results that shifting $\sigma_{\rm Ev}$ can shift the equilibrium oxygen vacancy concentration \cite{park_accurate_2023}. The analytical expression in Eq. \ref{eq:H_approx} gives an interpretable starting point for researchers to harness disorder-induced energy broadening to tune defect thermodynamics. 
\begin{equation}
    \Delta H^f \approx \hat{E}_{\rm v}  -  \frac{ \sigma_{\rm Ev}^2 }{k_{\rm B} T}
    \label{eq:H_approx}
\end{equation}
\begin{equation}
    \Delta S^f \approx \frac{1}{2}  S_{\rm O_2}  - \frac{\sigma_{\rm Ev}^2}{2 k_{\rm B} T^2} - S_{\rm O,vib}
    \label{eq:S_approx}
\end{equation}

The second derived expression, Eq. \ref{eq:S_approx}, is a departure from the common assumption $\Delta S^f = \frac{1}{2} S_{O_2}$ for oxygen vacancies \cite{park_accurate_2023, xu_local_2024, luo_predicting_2014}, and also describes the effects of oxygen vacancy energy broadening on $\Delta S^f$. Broadening $g(\hat{E}_{\rm v}, \sigma^2_{\rm Ev})$ always reduces the effective change in entropy. This is because introducing oxygen sites that favor vacancies (low $E_{v}$) increases the likelihood of configurations where those sites are vacant, lowering the configurational entropy. Skewing $p_{{\rm v},i}$ away from the uniform case, where $p_{{\rm v},i}=\hat{p}_{\rm v}$ always lowers configurational entropy. In short, having preferred sites for oxygen vacancies results in less configurational entropy, an effect propagated to a smaller $\Delta S^f$. This is similar to how high-entropy materials with preferred cation configurations have lower configurational entropy, reducing their phase stability \cite{sarker_high-entropy_2018, pitike_predicting_2020}. In Eq. \ref{eq:S_approx}, we also considered the phonon (vibrational) entropy contribution by an oxygen atom in the lattice, which was normally assumed to be small. Our estimated value of $S_{\rm O,vib}$ based on simulations described in the Supplementary Section 6 is not small, but about 25 J mol$^{-1}$ K$^{-1}$ (or 3R) for both LE and HE materials. 

We now have a theoretical basis for explaining the experimentally observed enthalpy-entropy correlation (compensation mechanism) for oxygen vacancies (Figure \ref{fig:empirical_panel}e). Eqs. \ref{eq:H_approx} and \ref{eq:S_approx}, found by approximating the first-order effects of $\sigma_{\rm Ev}$ from first-principles, correctly predict HE samples with higher $\sigma_{\rm Ev}$ will exhibit lower $\Delta H^f$ and $\Delta S^f$. In fact, Eqs. \ref{eq:H_approx} and \ref{eq:S_approx} can be combined in Eq. \ref{eq:HS_comp_approx} to predict a $\Delta H^f$ - $\Delta S^f$ scaling relation when $\hat{E_{\rm v} }$, $ S_{\rm O_2}$, and $S_{\rm O,vib}$ are constant. 

\begin{equation}
    \Delta H^f  = 2T\Delta S^f+ (\hat{E_{\rm v} } - T  S_{\rm O_2}+ 2TS_{\rm O,vib})
    \label{eq:HS_comp_approx}
\end{equation}

Eq. \ref{eq:HS_comp_approx} suggests the scaling between $\Delta H^f$ vs $\Delta S^f$ from perturbing $\sigma_{\rm Ev}$ is approximately $2T$. This prediction can be compared with our experimental results because simulated $\hat{E_{\rm v} }$ values are comparable between LE and HE samples (Table \ref{tab:sim_results}). The slope of the line of best fit in Figure \ref{fig:empirical_panel}e, known as the compensation or isoequilibrium temperature \cite{cornish-bowden_entropy-enthalpy_2018}, is found to be 1120K, a comparable order of magnitude to the predicted slope of $2T$ (1500-2500K). 

The scaling relation between $\Delta H^f$ and $\Delta S^f$ has opposing effects on the Gibb's of formation $\Delta G^f$ with the enthalpy term winning. For a small vacancy fraction $x$, broadening $g(\hat{E}_{\rm v}, \sigma^2_{\rm Ev})$ with always lower $\Delta G^f$ (Eq.  \ref{eq:G_approx}). 

\begin{equation}
    \Delta G^f  \approx \hat{E_{\rm v} } - T (\frac{1}{2}S_{\rm O_2} - S_{\rm O,vib}) -\frac{\sigma_{\rm Ev}^2}{2k_{\rm B} T}
    \label{eq:G_approx}
\end{equation}

Eq. \ref{eq:G_approx} extends to the equilibrium constant $K_p$ in Eq. \ref{eq:Kp_extended} , suggesting effects from $\sigma_{\rm Ev}$ could result in quadratic behavior that extends to the Van't Hoff diagram.

\begin{equation}
    \ln(K_p) = -\frac{\Delta G^f}{k_{\rm B}T} =
    \frac{\sigma_{\rm Ev}^2}{2k_{\rm B}^2 T^2}
    -\frac{ \hat{E_{\rm v} }}{k_{\rm B}T}
    +\frac{\frac{1}{2}\Delta S_{\rm O_2}-S_{\rm O,vib}}{k_{\rm B}}
    \label{eq:Kp_extended}
\end{equation}

In the following section, we evaluate the effectiveness of our theoretical and simulation approach by direct comparison with experimental data. 
\begin{figure*}
    \centering
    \includegraphics[width=1.0\linewidth]{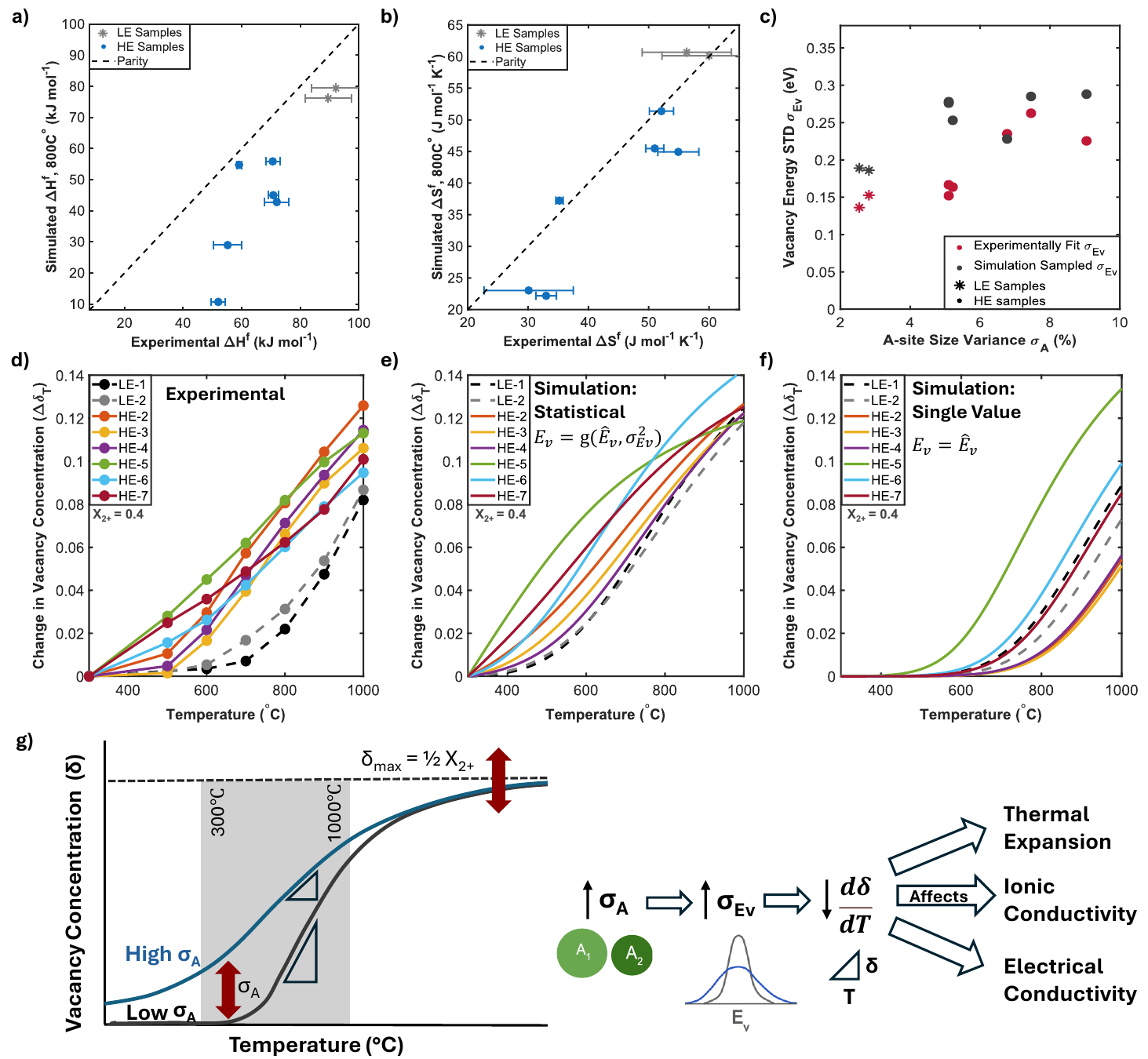}
    \caption{ \textbf{Comparison of simulation predictions and experiments. a} Experimental and simulated formation enthalpy ($\Delta H_f$) from Eq. \ref{eq:H_approx} with standard errors. \textbf{b} Experimental and simulated formation entropy ($\Delta S_f$) from Eq. \ref{eq:S_approx}. \textbf{c} Experimental variance in vacancy energy ($\sigma_{E_{\rm v}}$) found from Eq. \ref{eq:Kp_extended} compared to simulated $\sigma_{Ev}$. \textbf{d} Experimental changes in vacancy concentration ($\Delta \delta_T$) with temperature. \textbf{e} Model predictions where vacancy energies ($E_{\rm v}$) are treated as a distribution. \textbf{f} Model predictions where $E_{\rm v}$ is a single-valued average. \textbf{g} Schematic of effects of A-site size variance ($\sigma_A$) and 2+ site fraction ($X_{2+}$) on vacancy formation curves generated with Eq. \ref{eq:Kp_extended} for $\sigma_{A}$= 0 and 0.2 eV with experimental temperature regime shaded. Differing A-site cation sizes (green) increase $\sigma_{E_{\rm v}}$ affecting various bulk properties. Source data are provided as a Source Data file.} 
    \label{fig:compare_panel}
\end{figure*}

\subsection{Statistical Oxygen Vacancy Predictions}

We now compare the theoretically derived $\Delta H^f$ from Eq. \ref{eq:H_approx} and $\Delta S^f$  from Eq. \ref{eq:S_approx} with values extracted from experiments for all LE and HE samples (Figures \ref{fig:compare_panel}a and b). Since the experiments occur in temperature range of 500-1000$^\circ C$, we used an intermediate temperature of 800$^\circ C$ when using Eqs. \ref{eq:H_approx} and \ref{eq:S_approx}. We applied a Gaussian fit to $g(\hat{E}_{\rm v}, \sigma^2_{\rm Ev})$ sampled data in simulation (Table \ref{tab:sim_results}) to find values for $\hat{E}_{\rm v}$ and $\sigma_{\rm Ev}$, and used $S_{\rm O,vib}=25$ J $^{-1}$mol K$^{-1}$ (or $3k_{\rm B}$) estimated with simulations detailed in Supplementary Section 6. This value, averaged over four compositions, is similar to a previous DFT-calculated $S_{\rm O,vib}\approx2 k_{\rm B}$ for $\text{LaFeO}_3$ at 800$^\circ C$ \cite{baldassarri_vibrational_2024}. Simulations predict  $\Delta H^f$ values well (Figure \ref{fig:compare_panel}a) with a strong correlation of $R^2>0.8$. Calculated $\Delta S^f$ values (Figure \ref{fig:compare_panel}b) have good quantitative agreement with experimental values, especially for LE samples. Overall, Eqs. \ref{eq:H_approx} and \ref{eq:S_approx} exhibit good predictive power, while factors such as sensible enthalpy, vacancy-vacancy coupling effects, experimental uncertainty, vibrational entropy, and accuracy of the Matlantis potential could account for remaining discrepancies. In particular, our comparative analyses suggests vibrational entropy variance (Supplementary Section 4) and ML potential error with respect to DFT (Supplementary Section 6) could explain much the model error in Figures 4a-b,e.

To further validate the thermodynamically derived Eq. \ref{eq:Kp_extended}, we used this equation to extract the value of $\sigma_{\rm Ev}$ from experimental data and compared it with the simulated $\sigma_{\rm Ev}$ in Table \ref{tab:sim_results} (Figure \ref{fig:compare_panel}c). Specifically, we use experimentally-determined $\Delta \delta_T$ values, a tabulated value for $S_{O_2}$ \cite{janaf}, and $S_{\rm O,vib}$=25 J mol$^{-1}$ K$^{-1}$ (Supplementary Section 6) to fit values of $\sigma_{\rm Ev}$, $\hat{E}_{\rm v}$, and $\delta_0$. Figure \ref{fig:compare_panel}c shows $\sigma_{\rm Ev}$ values fit to \ref{eq:Kp_extended} closely match the simulation sampled values (Table \ref{tab:sim_results}) and both values of $\sigma_{\rm Ev}$ increase with increasing $\sigma_A$, indicating that the A-site size variation causes a vacancy formation energy distribution.

Finally, we compare experimentally measured oxygen vacancy concentration changes, $\Delta \delta_T$ (Figure \ref{fig:compare_panel}d), with respect to $\delta(300^\circ C)$,  with simulated values based on the current statistical approach with the full distribution of vacancy energies from simulations ($E_{\rm v} = g(\hat{E}_{\rm v}, \sigma^2_{\rm Ev})$),(Figure \ref{fig:compare_panel}e), and a classical average value ($E_{\rm v} = \hat{E}_{\rm v}$) approach (Figure \ref{fig:compare_panel}f). Experimental data in Figure \ref{fig:compare_panel}d are eight samples with ($X_{2+}$=0.4) from Figure \ref{fig:tga_panel}a. For both approaches, we used sampled values of $E_{\rm v}$ from the simulation in coordination with the unsimplified Eqs. \ref{eq:grand_canonical}-\ref{eq:vac_from_prob}, assuming $S_{O_2}$=205 J mol$^{-1}$ K$^{-1}$ \cite{janaf} and $S_{\rm O,vib}$=25 J mol$^{-1}$ K$^{-1}$. We added additional terms for effects of $pO_2$ and electron hole charge coupling (Eq. \ref{eq:charge_neutrality}) with further details in the Supplementary Section 5. Oxygen vacancy concentrations are found by solving Supplementary Eq. 35 numerically. This model expands one the model suggested by Park et al. \cite{park_accurate_2023} by coupling to vacancy formation to polaron destruction, considering oxygen vibrational entropy ($S_{\rm O,vib}\neq0$), and increasing site sampling with a machine-learned interatomic potential. Figures \ref{fig:compare_panel}d-f clearly show that the statistical model better matches experimental values and sample trends. Low-temperature overestimates of LE samples (Figure \ref{fig:compare_panel}e) could be due to insufficient activation energy to achieve equilibrium. Importantly, the statistical model accurately captures the oxygen formation behavior vs. temperature in terms of the curvatures of LE and HE materials. For example, the model correctly predicts HE-5 and HE-7 to exhibit the largest vacancy increases at low temperatures, with HE-7 beginning to plateau at high temperatures due to polaron hole depletion when $\delta \rightarrow \frac{1}{2} X_{2+}$ (see Section \ref{sec:defect_modeling}). In contrast, the conventional oxygen vacancy prediction model (Figure \ref{fig:compare_panel}f) fails to correctly differentiate LE and HE sample behavior. 

This work provides the first broad experimental validation of the statistical vacancy model proposed by \textit{Park}~\cite{park_accurate_2023} and \textit{Xu}~\cite{xu_local_2024}, demonstrating that $g(\hat{E}_{\rm v}, \sigma^2_{\rm E_v})$ more accurately captures oxygen-vacancy behavior than a single mean $\hat{E}_{\rm v}$. Our results confirm that modeling variations between oxygen sites is crucial for predicting the unique behavior of oxygen vacancies in our high-entropy perovskite oxides. In addition, our theoretical analysis (Section~\ref{sec:theory}) provides physical insight into how vacancy-energy distributions influence oxygen vacancy formation dynamics. Finally, comparison across 14 A-site compositions identifies $\sigma_A$ as a key composition-property descriptor for guiding future HEPO design.

Figure \ref{fig:compare_panel}g schematically summarizes this paper's findings. Across the 14 materials tested, two compositional parameters appear to dominate oxygen vacancy behavior: mole fraction of A-site divalent cations $X_{2+}$ and A-site size variance $\sigma_A$. The A-site divalent cations mole fraction $X_{2+}$ sets an upper limit for oxygen vacancy concentration in oxidizing environments as it provides available charge \textcolor{black}{per Eq. \ref{eq:charge_neutrality} \cite{oishi_oxygen_2008, bae_investigations_2019}.} The A-site size variations $\sigma_A$, as a structural descriptor, affect the vacancy energy variations $\sigma_{\rm Ev}$, which have derivable consequences on the vacancy thermodynamics. A larger $\sigma_A$ increases $\sigma_{\rm Ev}$, which stretches the vacancy formation curve horizontally (on a $\delta-T$ plot) and results in a less temperature-dependent oxygen vacancy change (e.g., smaller slopes). This suggests that properties closely related to oxygen vacancy concentrations, such as thermal expansion, oxygen ionic conductivity, and electrical conductivity, will also have less temperature dependence with a larger $\sigma_A$. 

The above model has only been validated for perovskite oxides in air with  $\rm Fe_{0.8}Co_{0.2}$ B-sites  and divalent doping of 0.17 $\leq X_{2+} \leq$ 0.5, however, the theoretical framework behind energetically heterogeneous oxygen sites could be relevant to other complex oxides. 

\subsection{Discussion}

In summary, we investigated the dependence of oxygen vacancy concentration on temperature for eleven HE and three LE, LSCF-based perovskites through a combined experimental, computational, and theoretical workflow. Our experimental results show that the HE perovskites generally have higher concentrations of oxygen vacancy, but their concentration increases relatively more slowly than that of LE samples. When those experimental data were used to extract the oxygen vacancy formation $\Delta H^f$, $\Delta S^f$, and $ K_{p,ox}$, they show strong correlations with both the conventional divalent doping concentration $X_{2+}$ and the newly identified A-site size variance $\sigma_A$.

To better understand this result, we conducted a computational investigation using a machine learning interatomic potential and found that A-site size variations distort oxygen-B-site bonds, adding variance to the oxygen bonding energy. We further derived oxygen vacancy formation thermodynamics with a statistical treatment of oxygen vacancies, described by the vacancy energy mean $\hat{E}_{\rm v}$ and variance $\sigma_{\rm Ev}$. The statistical model enabled experimental estimates of $\sigma_{\rm Ev}$ which align with simulations. The statistical model also predicted vacancy concentrations that align remarkably well with experiments compared to a conventional model. These results show that the A-site divalent cations mole fraction $X_{2+}$ sets an upper limit for oxygen vacancy concentration, and the A-site size variation $\sigma_A$ affects the vacancy concentration dependence on temperature. A larger $\sigma_A$ increases $\sigma_{\rm Ev}$, which stretches the vacancy formation curve horizontally (on a $\delta-T$ plot) and results in a less temperature-dependent oxygen vacancy change (e.g., smaller slopes). 

Practically, those results suggest that the A-site size variance could be a simple knob to control properties closely related to oxygen vacancy concentrations, such as thermal expansion, oxygen ionic conductivity, and electrical conductivity, especially their temperature-dependent behaviors. Although this study focuses on A-site cation substitution in LSCF-based perovskite oxides in oxidizing conditions, similar strategies for tuning the variance in formation energy may be theoretically extended to control defect-related properties in other high entropy materials.

\section {Methods}
\footnotesize

\subsection{Composition Descriptor Calculations}
Materials reported are chosen to span low and high entropy variations of the common SOEC material LSCF (LE-1 in Table \ref{table:sample_descriptors}). All have the same B-site, (Co\textsubscript{0.2}Fe\textsubscript{0.8}), as LSCF. The B-site's Co-Fe ratio is known to govern electrical conductivity and thermal expansion and the chosen composition has been refined in the SOEC literature to strike a balance between the two properties \cite{jiang_development_2019}. Of 14 compositions chosen, 3x have low-entropy (LE) A-sites (2-3 cation elements) and 11x samples have high-entropy (HE) A-sites ($\geq$ 5 cation elements). Some A-site elements were chosen randomly as equimolar combinations of Lanthanides and Alkaline Earth (AE) metals, while others were chosen to span a set of common compositional parameters describing the set of cations in the A-site (Table \ref{table:sample_descriptors}). We selected several parameters to quantitatively describe the characteristics of the A sites. The first parameter is the A-site molar fraction of divalent AE metals (Ca, Sr, or Ba) $X_{2+}$ (Eq.\ref{eq:X_2_plus}) which is known to shift the charge balance relative to other trivalent (3+ oxidation) Lanthanide A-site cations. 

\begin{equation}
    X_{2+} = X_{Ca} + X_{Sr} + X_{Ba}
    \label{eq:X_2_plus}
\end{equation}

The Goldschmidt tolerance factor $t$ (Eq.\ref{eq:tolerance_factor}) \cite{goldschmidt_gesetze_1926} relates to the average A-site cation ionic radius ($\hat{r})$ while A-site size variance $\sigma_A$ (Eq.\ref{eq:size_variance} earlier in text) describes the variance among $N$ different A-site cation radii ($r_i$) respectively. 

\begin{equation}
    t = \frac{\hat{r}_A + r_O}{\sqrt{2}(\hat{r}_B + r_O)}
    \label{eq:tolerance_factor}
\end{equation}

The last parameter, $\Delta S_{mix}$, is the A-site entropy of mixing $N$ elements with mole fractions $X_i$ and is larger for more elements mixed. Table \ref{table:sample_descriptors} lists 14 materials than span a wide range of $X_{2+}$, $t$, $\sigma_A$, and $\Delta S_{mix}$  to clarify which parameters strongly affect the formation of oxygen vacancies. 

\begin{equation}
    \Delta S_{mix} = -R \sum_i^N X_i \ln X_i
    \label{eq:entropy_mixing}
\end{equation}

To calculate tolerance factors $t$ (Eq. \ref{eq:tolerance_factor}) and size variances $\sigma_A$ (Eq. \ref{eq:size_variance}), Shannon ionic radii values are used \cite{shannon_revised_1976}. It is assumed the coordination number of oxygen is two, B-sites is 6, and A-sites is 12. While the oxidation and spin states of iron and cobalt can be subject to change depending on the environment, it is assumed most iron and cobalt are in the 3+ oxidation state. Finally, we adopt assumptions about the spin states previously used for a similar material with iron in a high spin state and cobalt in an intermediate spin state taken as the average of the high and low spin state ionic radii \cite{gangopadhyay_understanding_2010}.

\subsection{Material Synthesis}
The solution combustion synthesis begins with stoichiometric mixing of metal nitrate precursors from Thermo Fisher Scientific: La(NO$_3$)$_3$ (99.9\%), Sr(NO$_3$)$_2$ ($\geq$99\%), Ca(NO$_3$)$_2$ ($\geq$99\%), Ba(NO$_3$)$_2$ ($\geq$99\%), Nd(NO$_3$)$_3\cdot$6(H$_2$O) (99.9\%), Sm(NO$_3$)$_3\cdot$6(H$_2$O) (99.9\%), Gd(NO$_3$)$_3\cdot$6(H$_2$O) (99.9\%), Y(NO$_3$)$_3\cdot$6(H$_2$O) (99.9\%), Fe(NO$_3$)$_2  \cdot$  9(H$_2$O) ($\geq$98\%), and Co(NO$_3$)$_2\cdot$6(H$_2$O) ($\geq$98\%). Precursors were dissolved in 100mL of deionized water to form a 1M solution of precursor cations in a 500mL Pyrex beaker. Sucrose (Sigma-Aldrich Omnipur) is added as a fuel in a fuel-oxidizer ratio of 1.0 (1M sucrose). The solution is stirred with a magnetic stir rod and dehydrated at 50$^\circ C$ on a hot plate for $>$12 hours or until solution forms a gel. The rod is removed and temperature raised to 350$^\circ C$ to initiate the combustion reaction. When the gel had fully burned, about 30 minutes, the product is removed and ground with a mortar and pestle. The powders then are placed in an alumina crucible and annealed at 800$^\circ C$ for 90 minutes with a 4$^\circ C$ min$^{-1}$ ramp rate. 
Powders collected from the solution combustion synthesis were processed with high-energy ball-milling. Oxide powders ($\sim$10g) were dissolved in \~30mL of hexane (99.9\% Fischer Chemical) and placed in a 50mL stainless steel vial in a Retsch Cryomill. The powder underwent 3 hours of room-temperature wet-milling at 30Hz. Powders were collected after drying the resulting slurry on a 50$^\circ C$ hot plate. 

\subsection{Characterization Experiments}
The crystal structure (see Supplementary Figure 1) of the synthesized samples was investigated by X-ray diffraction (XRD, PANalytical Empyrean) with a Cu source (K$\alpha$,
$\qty{1.54}{\angstrom}$ of the wavelength).
The morphologies of synthesized powders were examined using scanning electron microscopy (SEM) on an FEI Magellan 400 XHR, operating at an accelerating voltage of 5 kV and a beam current of 25 pA. SEM/EDS was conducted using the same instrument with an accelerating voltage of 20 kV and a beam current of 3.2 nA. 

TGA is a common approach to measuring oxygen vacancies \cite{oishi_oxygen_2008, luo_predicting_2014, choi_thermodynamic_2012, bae_investigations_2019, mizusaki_nonstoichiometry_1985} and was found to be in agreement with iodometric titration and columbic titration methods \cite{karppinen_oxygen_2002}. The TGA protocol starts with a 1.5 hour 900$^\circ$C preheating step to ensure mass stability. The 50-70mg samples are cooled and reheated with 1-2 hour holds every 100$^\circ$C from 500$^\circ$C to 1000$^\circ$C in a Setaram LABSYS evo TGA-DSC (see Supplementary Figure 2). Experiments were performed in air ($pO_2$=0.21 atm) to be consistent with high $pO_2$ present for SOEC air electrode. TGA mass changes at each temperature $\Delta m_T$ (mg) relative to the initial mass $m_0$ (mg) were used to calculate the oxygen non-stoichiometry changes $\Delta \delta_T$ for a material with a theoretical molar mass $M_w$ with Eq.~\ref{eq:convert_TGA_delta}. 

\begin{equation}
    \Delta\delta_T = \frac{\Delta m_T}{m_0}\frac{M_w}{16}
    \label{eq:convert_TGA_delta}
\end{equation}

\subsection{Atomistic Simulations}
In this paper, the Matlantis PFP estimator v7.0.0, trained on data without the Hubbard U correction, is used to perform structural relaxation using the LBFGS algorithm. We conducted atomistic simulations using an MLUIP provided by Matlantis \cite{Matlantis}. Its stated mean average error of 0.03 eV for disordered systems is small compared to calculated vacancy formation energy values in the range of 0.15-1.4 eV \cite{matlantis_paper}. The structural analysis determined B-site bonding to oxygen with $\qty{1.1}{\angstrom}$  radius collision spheres and used a relaxation force tolerance of $f<\qty{0.05}{eV \angstrom^{-1}}$. For vacancy formation energy, the program iterates over removing 768 oxygen sites in a 1280 atom supercell (Supplementary Figure 3) randomly populated with cations in stoichiometric ratios with no initial oxygen vacancies.

\section{Data Availability}
Processed TGA vacancy concentration data, defect model fit paremters, and simulation calculated vacancy energies have been deposited in a FigShare database (DOI \hyperlink{https://doi.org/10.6084/m9.figshare.30015193}{10.6084/m9.figshare.30015193}). In addition, initial and relaxed supercell structures used for simulations are available on FigShare. Figure source data are provided as a Source Data file.

\section{Code Availability}
All code for atomic simulations is available on CodeOcean (DOI \hyperlink{https://doi.org/10.24433/CO.5451446.v1}{10.24433/CO.5451446.v1}), edited to allow for alternative ASE-compatible potentials.

\section*{Acknowledgment}
The authors gratefully acknowledge the financial and technical support provided by Genvia and the Commissariat à l'Énergie Atomique et aux Énergies Alternatives (CEA), France. We also acknowledge  support from the Stanford energy postdoctoral fellowship.

\section{Author Contributions}
A.P. lead experimental and simulation investigations, conceptualization, formal analysis (equal with J.Q.), and supplied the initial draft. Y.W. supported conceptualization, methodology, and writing review and editing. K.H. performed DFT validation studies. D.K. and Y.L. performed additional experiments and supported the methodology development. J.Q. performed formal analysis and provided supervision. X.Z.  lead project supervision, funding acquisition, and  writing review and editing.

\section{Competing Interests}
The authors declare no competing interests.


\begin{thebibliography}{10}
\expandafter\ifx\csname url\endcsname\relax
  \def\url#1{\texttt{#1}}\fi
\expandafter\ifx\csname urlprefix\endcsname\relax\def\urlprefix{URL }\fi
\providecommand{\bibinfo}[2]{#2}
\providecommand{\eprint}[2][]{\url{#2}}

\bibitem{pikalova_high-entropy_2022}
\bibinfo{author}{Pikalova, E.~Y.}, \bibinfo{author}{Kalinina, E.~G.}, \bibinfo{author}{Pikalova, N.~S.} \& \bibinfo{author}{Filonova, E.~A.}
\newblock \bibinfo{title}{High-entropy materials in {SOFC} technology: Theoretical foundations for their creation, features of synthesis, and recent achievements}.
\newblock \emph{\bibinfo{journal}{Materials}} \textbf{\bibinfo{volume}{15}}, \bibinfo{pages}{8783} (\bibinfo{year}{2022}).
\newblock \urlprefix\url{https://www.mdpi.com/1996-1944/15/24/8783}.
\newblock \bibinfo{note}{Number: 24 Publisher: Multidisciplinary Digital Publishing Institute}.

\bibitem{xiang_high-entropy_2021}
\bibinfo{author}{Xiang, H.} \emph{et~al.}
\newblock \bibinfo{title}{High-entropy ceramics: Present status, challenges, and a look forward}.
\newblock \emph{\bibinfo{journal}{Journal of Advanced Ceramics}} \textbf{\bibinfo{volume}{10}}, \bibinfo{pages}{385--441} (\bibinfo{year}{2021}).
\newblock \urlprefix\url{https://link.springer.com/10.1007/s40145-021-0477-y}.

\bibitem{bae_defect_2020}
\bibinfo{author}{Bae, H.}, \bibinfo{author}{Kim, I.-H.}, \bibinfo{author}{Singh, B.}, \bibinfo{author}{Bhardwaj, A.} \& \bibinfo{author}{Song, S.-J.}
\newblock \bibinfo{title}{Defect chemistry of highly defective la0.1sr0.9co0.8fe0.2o3-$\delta$ by considering oxygen interstitials: Effect of hole degeneracy}.
\newblock \emph{\bibinfo{journal}{Solid State Ionics}} \textbf{\bibinfo{volume}{347}}, \bibinfo{pages}{115251} (\bibinfo{year}{2020}).
\newblock \urlprefix\url{https://linkinghub.elsevier.com/retrieve/pii/S0167273819311725}.

\bibitem{choi_thermodynamic_2012}
\bibinfo{author}{Choi, M.-B.}, \bibinfo{author}{Jeon, S.-Y.}, \bibinfo{author}{Im, H.-N.} \& \bibinfo{author}{Song, S.-J.}
\newblock \bibinfo{title}{Thermodynamic quantities and oxygen nonstoichiometry of undoped {BaTiO}3- by thermogravimetric analysis}.
\newblock \emph{\bibinfo{journal}{Journal of Alloys and Compounds}} \textbf{\bibinfo{volume}{513}}, \bibinfo{pages}{487--494} (\bibinfo{year}{2012}).
\newblock \urlprefix\url{https://linkinghub.elsevier.com/retrieve/pii/S0925838811020652}.

\bibitem{rost_entropy-stabilized_2015}
\bibinfo{author}{Rost, C.~M.} \emph{et~al.}
\newblock \bibinfo{title}{Entropy-stabilized oxides}.
\newblock \emph{\bibinfo{journal}{Nature Communications}} \textbf{\bibinfo{volume}{6}}, \bibinfo{pages}{8485} (\bibinfo{year}{2015}).
\newblock \urlprefix\url{https://www.nature.com/articles/ncomms9485}.
\newblock \bibinfo{note}{Publisher: Nature Publishing Group}.

\bibitem{su_direct_2022}
\bibinfo{author}{Su, L.} \emph{et~al.}
\newblock \bibinfo{title}{Direct observation of elemental fluctuation and oxygen octahedral distortion-dependent charge distribution in high entropy oxides}.
\newblock \emph{\bibinfo{journal}{Nature Communications}} \textbf{\bibinfo{volume}{13}}, \bibinfo{pages}{2358} (\bibinfo{year}{2022}).
\newblock \urlprefix\url{https://www.nature.com/articles/s41467-022-30018-y}.

\bibitem{xu_local_2024}
\bibinfo{author}{Xu, B.} \emph{et~al.}
\newblock \bibinfo{title}{Local ordering, distortion, and redox activity in (la0.75sr0.25)(mn0.25fe0.25co0.25al0.25)o3 investigated by a computational workflow for compositionally complex perovskite oxides}.
\newblock \emph{\bibinfo{journal}{Chemistry of Materials}} \textbf{\bibinfo{volume}{36}}, \bibinfo{pages}{4990--5001} (\bibinfo{year}{2024}).
\newblock \urlprefix\url{https://doi.org/10.1021/acs.chemmater.3c03038}.
\newblock \bibinfo{note}{Publisher: American Chemical Society}.

\bibitem{zhang_tuning_2023}
\bibinfo{author}{Zhang, M.} \emph{et~al.}
\newblock \bibinfo{title}{Tuning oxygen vacancies in oxides by configurational entropy}.
\newblock \emph{\bibinfo{journal}{{ACS} Applied Materials \& Interfaces}} \textbf{\bibinfo{volume}{15}}, \bibinfo{pages}{45774--45789} (\bibinfo{year}{2023}).
\newblock \urlprefix\url{https://doi.org/10.1021/acsami.3c07268}.
\newblock \bibinfo{note}{Publisher: American Chemical Society}.

\bibitem{choi_oxygen_2012}
\bibinfo{author}{Choi, M.-B.}, \bibinfo{author}{Lim, D.-K.}, \bibinfo{author}{Wachsman, E.} \& \bibinfo{author}{Song, S.-J.}
\newblock \bibinfo{title}{Oxygen nonstoichiometry and chemical expansion of mixed conducting la0.1sr0.9co0.8fe0.2o3-$\delta$}.
\newblock \emph{\bibinfo{journal}{Solid State Ionics}} \textbf{\bibinfo{volume}{221}}, \bibinfo{pages}{22--27} (\bibinfo{year}{2012}).
\newblock \urlprefix\url{https://linkinghub.elsevier.com/retrieve/pii/S016727381200375X}.

\bibitem{oishi_oxygen_2008}
\bibinfo{author}{Oishi, M.}, \bibinfo{author}{Yashiro, K.}, \bibinfo{author}{Sato, K.}, \bibinfo{author}{Mizusaki, J.} \& \bibinfo{author}{Kawada, T.}
\newblock \bibinfo{title}{Oxygen nonstoichiometry and defect structure analysis of b-site mixed perovskite-type oxide (la, sr)(cr, m)o3-$\delta$ (m=ti, mn and fe)}.
\newblock \emph{\bibinfo{journal}{Journal of Solid State Chemistry}} \textbf{\bibinfo{volume}{181}}, \bibinfo{pages}{3177--3184} (\bibinfo{year}{2008}).
\newblock \urlprefix\url{https://linkinghub.elsevier.com/retrieve/pii/S0022459608004477}.

\bibitem{luo_predicting_2014}
\bibinfo{author}{Luo, H.} \emph{et~al.}
\newblock \bibinfo{title}{Predicting oxygen vacancy non-stoichiometric concentration in perovskites from first principles}.
\newblock \emph{\bibinfo{journal}{Applied Surface Science}} \textbf{\bibinfo{volume}{323}}, \bibinfo{pages}{65--70} (\bibinfo{year}{2014}).
\newblock \urlprefix\url{https://linkinghub.elsevier.com/retrieve/pii/S0169433214014007}.

\bibitem{bae_investigations_2019}
\bibinfo{author}{Bae, H.} \emph{et~al.}
\newblock \bibinfo{title}{Investigations on defect equilibrium, thermodynamic quantities, and transport properties of la$_{\textrm{0.5}}$ sr$_{\textrm{0.5}}$ {FeO}$_{\textrm{3}-\delta}$}.
\newblock \emph{\bibinfo{journal}{Journal of The Electrochemical Society}} \textbf{\bibinfo{volume}{166}}, \bibinfo{pages}{F180--F189} (\bibinfo{year}{2019}).
\newblock \urlprefix\url{https://iopscience.iop.org/article/10.1149/2.0311904jes}.

\bibitem{park_accurate_2023}
\bibinfo{author}{Park, J.} \emph{et~al.}
\newblock \bibinfo{title}{Accurate prediction of oxygen vacancy concentration with disordered a-site cations in high-entropy perovskite oxides}.
\newblock \emph{\bibinfo{journal}{npj Computational Materials}} \textbf{\bibinfo{volume}{9}}, \bibinfo{pages}{29} (\bibinfo{year}{2023}).
\newblock \urlprefix\url{https://www.nature.com/articles/s41524-023-00981-1}.

\bibitem{mizusaki_nonstoichiometry_1985}
\bibinfo{author}{Mizusaki, J.}, \bibinfo{author}{Yoshihiro, M.}, \bibinfo{author}{Yamauchi, S.} \& \bibinfo{author}{Fueki, K.}
\newblock \bibinfo{title}{Nonstoichiometry and defect structure of the perovskite-type oxides la1-{xSrxFeO}3-$\delta$}.
\newblock \emph{\bibinfo{journal}{Journal of Solid State Chemistry}} \textbf{\bibinfo{volume}{58}}, \bibinfo{pages}{257--266} (\bibinfo{year}{1985}).
\newblock \urlprefix\url{https://www.sciencedirect.com/science/article/pii/0022459685902439}.

\bibitem{jeon_oxygen_2012}
\bibinfo{author}{Jeon, S.-Y.}, \bibinfo{author}{Choi, M.-B.}, \bibinfo{author}{Hwang, J.-H.}, \bibinfo{author}{Wachsman, E.~D.} \& \bibinfo{author}{Song, S.-J.}
\newblock \bibinfo{title}{Oxygen excess nonstoichiometry and thermodynamic quantities of la2nio4 + $\delta$}.
\newblock \emph{\bibinfo{journal}{Journal of Solid State Electrochemistry}} \textbf{\bibinfo{volume}{16}}, \bibinfo{pages}{785--793} (\bibinfo{year}{2012}).
\newblock \urlprefix\url{http://link.springer.com/10.1007/s10008-011-1427-3}.

\bibitem{choi_correlation_2014}
\bibinfo{author}{Choi, M.-B.} \emph{et~al.}
\newblock \bibinfo{title}{Correlation between defect structure and electrochemical properties of mixed conducting la0.1sr0.9co0.8fe0.2o3-}.
\newblock \emph{\bibinfo{journal}{Acta Materialia}} \textbf{\bibinfo{volume}{65}}, \bibinfo{pages}{373--382} (\bibinfo{year}{2014}).
\newblock \urlprefix\url{https://linkinghub.elsevier.com/retrieve/pii/S1359645413008562}.

\bibitem{karppinen_oxygen_2002}
\bibinfo{author}{Karppinen, M.}, \bibinfo{author}{Matvejeff, M.}, \bibinfo{author}{Salomäki, K.} \& \bibinfo{author}{Yamauchi, H.}
\newblock \bibinfo{title}{Oxygen content analysis of functional perovskite-derived cobalt oxides}.
\newblock \emph{\bibinfo{journal}{Journal of Materials Chemistry}} \textbf{\bibinfo{volume}{12}}, \bibinfo{pages}{1761--1764} (\bibinfo{year}{2002}).
\newblock \urlprefix\url{https://xlink.rsc.org/?DOI=b200770n}.

\bibitem{deml_intrinsic_2015}
\bibinfo{author}{Deml, A.~M.}, \bibinfo{author}{Holder, A.~M.}, \bibinfo{author}{O’Hayre, R.~P.}, \bibinfo{author}{Musgrave, C.~B.} \& \bibinfo{author}{Stevanović, V.}
\newblock \bibinfo{title}{Intrinsic material properties dictating oxygen vacancy formation energetics in metal oxides}.
\newblock \emph{\bibinfo{journal}{The Journal of Physical Chemistry Letters}} \textbf{\bibinfo{volume}{6}}, \bibinfo{pages}{1948--1953} (\bibinfo{year}{2015}).
\newblock \urlprefix\url{https://pubs.acs.org/doi/10.1021/acs.jpclett.5b00710}.

\bibitem{jiang_development_2019}
\bibinfo{author}{Jiang, S.~P.}
\newblock \bibinfo{title}{Development of lanthanum strontium cobalt ferrite perovskite electrodes of solid oxide fuel cells – a review}.
\newblock \emph{\bibinfo{journal}{International Journal of Hydrogen Energy}} \textbf{\bibinfo{volume}{44}}, \bibinfo{pages}{7448--7493} (\bibinfo{year}{2019}).
\newblock \urlprefix\url{https://www.sciencedirect.com/science/article/pii/S0360319919303714}.

\bibitem{sivak_discovering_2025}
\bibinfo{author}{Sivak, J.~T.} \emph{et~al.}
\newblock \bibinfo{title}{Discovering high-entropy oxides with a machine-learning interatomic potential}.
\newblock \emph{\bibinfo{journal}{Physical Review Letters}} \textbf{\bibinfo{volume}{134}}, \bibinfo{pages}{216101} (\bibinfo{year}{2025}).
\newblock \urlprefix\url{https://link.aps.org/doi/10.1103/PhysRevLett.134.216101}.
\newblock \bibinfo{note}{Publisher: American Physical Society}.

\bibitem{pitike_predicting_2020}
\bibinfo{author}{Pitike, K.~C.}, \bibinfo{author}{{KC}, S.}, \bibinfo{author}{Eisenbach, M.}, \bibinfo{author}{Bridges, C.~A.} \& \bibinfo{author}{Cooper, V.~R.}
\newblock \bibinfo{title}{Predicting the phase stability of multicomponent high-entropy compounds}.
\newblock \emph{\bibinfo{journal}{Chemistry of Materials}} \textbf{\bibinfo{volume}{32}}, \bibinfo{pages}{7507--7515} (\bibinfo{year}{2020}).
\newblock \urlprefix\url{https://doi.org/10.1021/acs.chemmater.0c02702}.
\newblock \bibinfo{note}{Publisher: American Chemical Society}.

\bibitem{sarker_high-entropy_2018}
\bibinfo{author}{Sarker, P.} \emph{et~al.}
\newblock \bibinfo{title}{High-entropy high-hardness metal carbides discovered by entropy descriptors}.
\newblock \emph{\bibinfo{journal}{Nature Communications}} \textbf{\bibinfo{volume}{9}}, \bibinfo{pages}{4980} (\bibinfo{year}{2018}).
\newblock \urlprefix\url{https://www.nature.com/articles/s41467-018-07160-7}.
\newblock \bibinfo{note}{Publisher: Nature Publishing Group}.

\bibitem{kuhn_oxygen_2011}
\bibinfo{author}{Kuhn, M.}, \bibinfo{author}{Hashimoto, S.}, \bibinfo{author}{Sato, K.}, \bibinfo{author}{Yashiro, K.} \& \bibinfo{author}{Mizusaki, J.}
\newblock \bibinfo{title}{Oxygen nonstoichiometry, thermo-chemical stability and lattice expansion of la0.6sr0.4feo3-$\delta$}.
\newblock \emph{\bibinfo{journal}{Solid State Ionics}} \textbf{\bibinfo{volume}{195}}, \bibinfo{pages}{7--15} (\bibinfo{year}{2011}).
\newblock \urlprefix\url{https://www.sciencedirect.com/science/article/pii/S0167273811002724}.

\bibitem{sogaard_oxygen_2007}
\bibinfo{author}{Søgaard, M.}, \bibinfo{author}{Vang~Hendriksen, P.} \& \bibinfo{author}{Mogensen, M.}
\newblock \bibinfo{title}{Oxygen nonstoichiometry and transport properties of strontium substituted lanthanum ferrite}.
\newblock \emph{\bibinfo{journal}{Journal of Solid State Chemistry}} \textbf{\bibinfo{volume}{180}}, \bibinfo{pages}{1489--1503} (\bibinfo{year}{2007}).
\newblock \urlprefix\url{https://www.sciencedirect.com/science/article/pii/S0022459607000862}.

\bibitem{hartmann_investigation_2023}
\bibinfo{author}{Hartmann, C.}, \bibinfo{author}{Geneste, G.}, \bibinfo{author}{Saravanabavan, K.} \& \bibinfo{author}{Laurencin, J.}
\newblock \bibinfo{title}{Investigation of the reaction mechanisms in la1-{XSrxCo}1-{YFeyO}3 electrode by {DFT} calculations}.
\newblock \emph{\bibinfo{journal}{{ECS} Transactions}} \textbf{\bibinfo{volume}{111}}, \bibinfo{pages}{1239} (\bibinfo{year}{2023}).
\newblock \urlprefix\url{https://iopscience.iop.org/article/10.1149/11106.1239ecst/meta}.
\newblock \bibinfo{note}{Publisher: {IOP} Publishing}.

\bibitem{starikov_enthalpyentropy_2007}
\bibinfo{author}{Starikov, E.~B.} \& \bibinfo{author}{Nordén, B.}
\newblock \bibinfo{title}{Enthalpy−entropy compensation: A phantom or something useful?}
\newblock \emph{\bibinfo{journal}{The Journal of Physical Chemistry B}} \textbf{\bibinfo{volume}{111}}, \bibinfo{pages}{14431--14435} (\bibinfo{year}{2007}).
\newblock \urlprefix\url{https://doi.org/10.1021/jp075784i}.
\newblock \bibinfo{note}{Publisher: American Chemical Society}.

\bibitem{cornish-bowden_entropy-enthalpy_2018}
\bibinfo{author}{Cornish-Bowden, A.}
\newblock \emph{\bibinfo{title}{Entropy-Enthalpy Compensation}} (\bibinfo{publisher}{Springer}, \bibinfo{year}{2018}).
\newblock \urlprefix\url{https://doi.org/10.1007/978-3-642-35943-9_10072-1}.

\bibitem{pan_enthalpyentropy_2016}
\bibinfo{author}{Pan, A.}, \bibinfo{author}{Kar, T.}, \bibinfo{author}{Rakshit, A.~K.} \& \bibinfo{author}{Moulik, S.~P.}
\newblock \bibinfo{title}{Enthalpy–entropy compensation ({EEC}) effect: Decisive role of free energy}.
\newblock \emph{\bibinfo{journal}{The Journal of Physical Chemistry B}} \textbf{\bibinfo{volume}{120}}, \bibinfo{pages}{10531--10539} (\bibinfo{year}{2016}).
\newblock \urlprefix\url{https://doi.org/10.1021/acs.jpcb.6b05890}.
\newblock \bibinfo{note}{Publisher: American Chemical Society}.

\bibitem{matlantis_paper}
\bibinfo{author}{Takamoto, S.} \emph{et~al.}
\newblock \bibinfo{title}{Towards universal neural network potential for material discovery applicable to arbitrary combination of 45 elements}.
\newblock \emph{\bibinfo{journal}{Nature Communications}} \textbf{\bibinfo{volume}{13}}, \bibinfo{pages}{2991} (\bibinfo{year}{2022}).
\newblock \urlprefix\url{https://doi.org/10.1038/s41467-022-30687-9}.

\bibitem{ase}
\bibinfo{author}{Larsen, A.~H.} \emph{et~al.}
\newblock \bibinfo{title}{The atomic simulation environment—a python library for working with atoms}.
\newblock \emph{\bibinfo{journal}{Journal of Physics: Condensed Matter}} \textbf{\bibinfo{volume}{29}}, \bibinfo{pages}{273002} (\bibinfo{year}{2017}).
\newblock \urlprefix\url{https://dx.doi.org/10.1088/1361-648X/aa680e}.
\newblock \bibinfo{note}{Publisher: {IOP} Publishing}.

\bibitem{jing_role_2020}
\bibinfo{author}{Jing, Y.} \& \bibinfo{author}{Aluru, N.}
\newblock \bibinfo{title}{The role of a-site ion on proton diffusion in perovskite oxides ({ABO}3)}.
\newblock \emph{\bibinfo{journal}{Journal of Power Sources}} \textbf{\bibinfo{volume}{445}}, \bibinfo{pages}{227327} (\bibinfo{year}{2020}).
\newblock \urlprefix\url{https://linkinghub.elsevier.com/retrieve/pii/S0378775319313205}.

\bibitem{zou_high-entropy-induced_2025}
\bibinfo{author}{Zou, G.} \emph{et~al.}
\newblock \bibinfo{title}{High-entropy-induced {CoO} 6 octahedral distortion for boosted oxygen evolution reaction at high temperature}.
\newblock \emph{\bibinfo{journal}{Energy \& Environmental Science}} \textbf{\bibinfo{volume}{18}}, \bibinfo{pages}{9478--9489} (\bibinfo{year}{2025}).
\newblock \urlprefix\url{https://pubs.rsc.org/en/content/articlelanding/2025/ee/d5ee01370d}.
\newblock \bibinfo{note}{Publisher: Royal Society of Chemistry}.

\bibitem{stevanovic_correcting_2012}
\bibinfo{author}{Stevanović, V.}, \bibinfo{author}{Lany, S.}, \bibinfo{author}{Zhang, X.} \& \bibinfo{author}{Zunger, A.}
\newblock \bibinfo{title}{Correcting density functional theory for accurate predictions of compound enthalpies of formation: Fitted elemental-phase reference energies}.
\newblock \emph{\bibinfo{journal}{Physical Review B}} \textbf{\bibinfo{volume}{85}}, \bibinfo{pages}{115104} (\bibinfo{year}{2012}).
\newblock \urlprefix\url{https://link.aps.org/doi/10.1103/PhysRevB.85.115104}.

\bibitem{rudisill_standard_1992}
\bibinfo{author}{Rudisill, E.~N.} \& \bibinfo{author}{{LeVan}, M.~D.}
\newblock \bibinfo{title}{Standard states for the adsorbed-solution theory}.
\newblock \emph{\bibinfo{journal}{Chemical Engineering Science}} \textbf{\bibinfo{volume}{47}}, \bibinfo{pages}{1239--1245} (\bibinfo{year}{1992}).
\newblock \urlprefix\url{https://www.sciencedirect.com/science/article/pii/0009250992802458}.

\bibitem{elmutasim_evolution_2024}
\bibinfo{author}{Elmutasim, O.} \emph{et~al.}
\newblock \bibinfo{title}{Evolution of oxygen vacancy sites in ceria-based high-entropy oxides and their role in n2 activation}.
\newblock \emph{\bibinfo{journal}{{ACS} Applied Materials \& Interfaces}} \textbf{\bibinfo{volume}{16}}, \bibinfo{pages}{23038--23053} (\bibinfo{year}{2024}).
\newblock \urlprefix\url{https://doi.org/10.1021/acsami.3c16521}.
\newblock \bibinfo{note}{Publisher: American Chemical Society}.

\bibitem{baldassarri_vibrational_2024}
\bibinfo{author}{Baldassarri, B.}, \bibinfo{author}{He, J.} \& \bibinfo{author}{Wolverton, C.}
\newblock \bibinfo{title}{Vibrational entropies of oxygen vacancy formation in metal oxides}.
\newblock \emph{\bibinfo{journal}{Physical Review Materials}} \textbf{\bibinfo{volume}{8}}, \bibinfo{pages}{055407} (\bibinfo{year}{2024}).
\newblock \urlprefix\url{https://link.aps.org/doi/10.1103/PhysRevMaterials.8.055407}.
\newblock \bibinfo{note}{Publisher: American Physical Society}.

\bibitem{janaf}
\bibinfo{author}{Chase, M.}
\newblock \emph{\bibinfo{title}{NIST-JANAF Thermochemical Tables, 4th Edition}} (\bibinfo{publisher}{American Institute of Physics, -1}, \bibinfo{year}{1998}).

\bibitem{goldschmidt_gesetze_1926}
\bibinfo{author}{Goldschmidt, V.~M.}
\newblock \bibinfo{title}{Die gesetze der krystallochemie}.
\newblock \emph{\bibinfo{journal}{Naturwissenschaften}} \textbf{\bibinfo{volume}{14}}, \bibinfo{pages}{477--485} (\bibinfo{year}{1926}).
\newblock \urlprefix\url{https://doi.org/10.1007/BF01507527}.

\bibitem{shannon_revised_1976}
\bibinfo{author}{Shannon, R.~D.}
\newblock \bibinfo{title}{Revised effective ionic radii and systematic studies of interatomic distances in halides and chalcogenides}.
\newblock \emph{\bibinfo{journal}{Acta Crystallographica Section A: Crystal Physics, Diffraction, Theoretical and General Crystallography}} \textbf{\bibinfo{volume}{32}}, \bibinfo{pages}{751--767} (\bibinfo{year}{1976}).
\newblock \urlprefix\url{https://journals.iucr.org/a/issues/1976/05/00/a12967/}.
\newblock \bibinfo{note}{Publisher: International Union of Crystallography}.

\bibitem{gangopadhyay_understanding_2010}
\bibinfo{author}{Gangopadhyay, S.} \emph{et~al.}
\newblock \bibinfo{title}{Understanding oxygen vacancy migration and clustering in barium strontiumcobalt iron oxide}.
\newblock \emph{\bibinfo{journal}{Solid State Ionics}} \textbf{\bibinfo{volume}{181}}, \bibinfo{pages}{1067--1073} (\bibinfo{year}{2010}).
\newblock \urlprefix\url{https://www.sciencedirect.com/science/article/pii/S0167273810002481}.

\bibitem{Matlantis}
\bibinfo{title}{{Matlantis, software as a service style material discovery tool}}.
\newblock \bibinfo{howpublished}{\url{https://matlantis.com/}} (\bibinfo{year}{2024}).

\end{thebibliography}

\end{document}